\documentclass[11pt]{article}
\usepackage{setspace}
\usepackage{changepage}
\usepackage{url}
\usepackage[left=1.5in,right=1.5in]{geometry}
\usepackage{float}
\usepackage{multirow}
\usepackage{color,xcolor}
\usepackage{listings}
\usepackage{caption}
\usepackage{subcaption} 
\usepackage{algorithmicx}
\usepackage[ruled]{algorithm}
\usepackage{algpseudocode}
\algnewcommand{\LeftComment}[1]{\State // #1}

\usepackage[
  linecolor=black,%
  innerleftmargin=0,
  innerrightmargin=6,
  skipabove=\topskip,
  innertopmargin=-9,
  usetwoside=false,
  linewidth=1pt
 ]{mdframed}
\DeclareCaptionFont{white}{\color{white}}
\DeclareCaptionFormat{listing}{%
 \parbox{\textwidth}{\colorbox{black}{\parbox{\textwidth}{#1#2#3}}\vskip3pt}}
\usepackage{graphicx}
\usepackage{hyperref}

\newcommand{\ThesisTitle}{A MapReduce Approach to NoSQL RDF Databases}

\usepackage{qtree} 
\def\ojoin{\setbox0=\hbox{$\bowtie$}%
  \rule[-.02ex]{.25em}{.4pt}\llap{\rule[\ht0]{.25em}{.4pt}}}
\def\leftouterjoin{\mathbin{\ojoin\mkern-5.8mu\bowtie}}

\begin{document}

\thispagestyle{empty}
\pagestyle{empty}
\pagenumbering{roman}

\newcommand{\SignatureField}[1]{
	\begin{tabular}{l}
		\rule{80 mm}{1pt} \\
		#1
	\end{tabular}
	\vspace{1cm}
}

\newpage
\doublespacing
\null\vfill
\begin{center}
	\LARGE
	\onehalfspacing
	\textbf{\ThesisTitle \\
		\doublespacing
		\Large
		\vspace{8 mm}
		by \\
		\vspace{8 mm}
		Albert Haque} \\
	\vspace{30 mm}
	\large\textbf{HONORS THESIS} \\
	\vspace{4 mm}
	\doublespacing\normalsize
	Presented to the Faculty of the Department of Computer Science \\
	The University of Texas at Austin \\
	in Partial Fulfillment \\
	of the Requirements \\
	for the Degree of \\
	\vspace{5 mm}
	\large
	\textbf{BACHELOR OF SCIENCE}\\
	\vspace{25 mm}
	THE UNIVERSITY OF TEXAS AT AUSTIN \\
	December 2013
\end{center}
\vfill

\newpage
\begin{center}
	\null\vfill
	\onehalfspacing
	\textbf{The Thesis Committee for Albert Haque \\
		Certifies that this is the approved version of the following thesis: \\
		\vspace{35 mm}
		\LARGE
		\ThesisTitle \\
	}
	\singlespacing
\end{center}
\normalsize
\vspace{33 mm}
\begin{adjustwidth}{57 mm}{0 mm}
	\textbf{Approved By \\
		Supervising Committee:
	}
\end{adjustwidth}
\singlespacing
\vspace{5 mm}
\begin{adjustwidth}{55 mm}{0 mm}
	\SignatureField{Daniel Miranker, Supervisor} \\
	\SignatureField{Lorenzo Alvisi} \\
	\SignatureField{Adam Klivans}
\end{adjustwidth}

\setcounter{secnumdepth}{-2}

\newpage
\pagestyle{plain}
\phantomsection
\addcontentsline{toc}{section}{Acknowledgments}
\doublespacing
\begin{center}
	\Large
	\textbf{Acknowledgments} \\
	\vspace{7 mm}
\end{center}
\normalsize
\par

First and foremost, I would like to thank Daniel Miranker, my thesis advisor, for his insight and guidance throughout the process of working on this thesis. His mentorship and encouragement were essential in developing this project from idea generation to maturity. I would like to thank all the members of the Research in Bioinformatics and Semantic Web lab for their encouragement and scholarly discussions over the years. I'd also like to thank my committee members, Lorenzo Alvisi and Adam Klivans, for their detailed feedback and for challenging the assumptions and logic I made in my research. Finally, I'd like to thank my family and friends for their continuous support and encouragement.
Your motivation has propelled me forward during all my academic pursuits. This work was supported by the National Science Foundation under Grant No. IIS-1018554.

\newpage
\pagestyle{plain}
\phantomsection
\addcontentsline{toc}{section}{Abstract}
\doublespacing
\begin{center}
	\Large
	\textbf{Abstract} \\
	\vspace{7 mm}
\end{center}
\normalsize
\par
In recent years, the increased need to house and process large volumes of data has prompted the need for distributed storage and querying systems. The growth of machine-readable RDF triples has prompted both industry and academia to develop new database systems, called ``NoSQL," with characteristics that differ from classical databases. Many of these systems compromise ACID properties for increased horizontal scalability and data availability.
\par
This thesis concerns the development and evaluation of a NoSQL triplestore. Triplestores are database management systems central to emerging technologies such as the Semantic Web and linked data.  A triplestore comprises data storage using the RDF (resource description framework) data model and the execution of queries written in SPARQL. The triplestore developed here exploits an open-source stack comprising, Hadoop, HBase, and Hive. The evaluation spans several benchmarks, including the two most commonly used in triplestore evaluation, the Berlin SPARQL Benchmark, and the DBpedia benchmark, a query workload that operates an RDF representation of Wikipedia. Results reveal that the join algorithm used by the system plays a critical role in dictating query runtimes. Distributed graph databases must carefully optimize queries before generating MapReduce query plans as network traffic for large datasets can become prohibitive if the query is executed naively.

\newpage
\singlespacing
\setcounter{tocdepth}{3}
\cleardoublepage
\phantomsection
\addcontentsline{toc}{section}{Table of Contents}
\tableofcontents

\newpage
\phantomsection
\addcontentsline{toc}{section}{List of Figures}
\listoffigures
\newpage
\phantomsection
\addcontentsline{toc}{section}{List of Tables}
\listoftables
\newpage

\setcounter{secnumdepth}{3}

\pagenumbering{arabic}

\doublespacing

\section{Introduction}

Resource Description Framework (RDF) is a model, or description, of information stored on the Internet. Typically this information follows unstructured schemas, a variety of syntax notations, and differing data serialization formats. All RDF objects take the form of a subject-predicate-object expression and are commonly used for knowledge representation and graph datasets. As a result, automated software agents can store, exchange, and use this machine-readable information distributed throughout the Internet.


\begin{figure}[h]
	\centering
	\caption[Taxonomy of RDF Data Management]{Taxonomy of RDF Data Management \cite{sequeda2012ultrawrap}}
	\includegraphics[width=.6\textwidth]{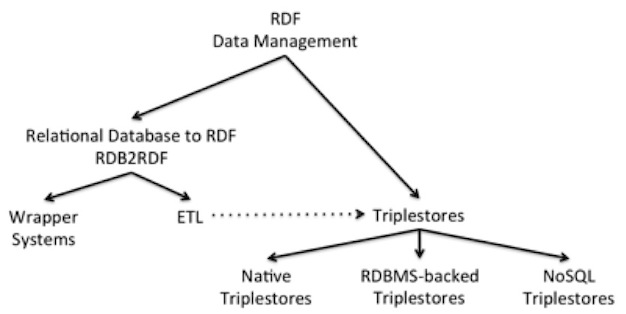}
	\label{fig:rdf_taxonomy}
\end{figure}

There are several methods for housing RDF data. We illustrate the taxonomy of these methods in Figure \ref{fig:rdf_taxonomy}. The traditional approach is to use relational database systems as triplestores. These systems have been the canonical form of data storage for several decades. Although these systems have good performance and query optimization techniques for most workloads, for large datasets, these systems are unable to feasibly scale to accommodate distributed operations. Attempts to store large amounts of RDF data on RDBMS have proven difficult \cite{ding2006characterizing}. As a result, alternatives to relational databases must be developed to satisfy the burgeoning need for ``big data" systems.

In recent years, an alternative class of database systems has emerged. NoSQL systems, which stand for ``not only SQL," try to solve the scalability limitations imposed by classical relational systems. Many NoSQL systems attempt to serve a large number of users simultaneously, efficiently scale to thousands of compute nodes, and trade full ACID properties for increased performance. Several major Internet companies such as Google \cite{dean2008mapreduce}, Facebook \cite{thusoo2009hive}, Microsoft\footnote{\url{http://www.windowsazure.com}}, and Amazon\footnote{\url{http://aws.amazon.com}} have deployed NoSQL systems in production environments.

In this thesis, we present a NoSQL triplestore used to store and query RDF data in a distributed fashion. Queries are executed using MapReduce and accept both SPARQL and SQL languages. This thesis is organized as follows: We begin with an overview of the technologies used by this system in Section \ref{sec:background}. In Section \ref{sec:data_layer}, we describe how RDF data is represented and stored in our distributed database system. Section \ref{sec:query_layer} addresses how we parse and transform SPARQL queries into MapReduce jobs. Sections \ref{sec:setup} and \ref{sec:results} describe our experimental setup, benchmarks, queries, and query performance followed by a discussion of these results in Section \ref{sec:discussion}. Finally we review related work in Section \ref{sec:related} before concluding in Section \ref{sec:conclusion}.

\section{Background} \label{sec:background}

\subsection{Semantic Web}

The Semantic Web was created to enable machines to automatically access and process a graph of linked data on the World Wide Web. Linked Data refers to structured data that typically obey following properties \cite{Berners-Lee:2006}:
(i) Uniform resource identifiers, or URIs, uniquely identify things, objects, or concepts.
(ii) Each URI points to an HTTP page displaying additional information about the thing or concept.
(iii) Established standards such as RDF and SPARQL are used to structure information.
(iv) Each concept or thing points to another concept or thing using a URI.

The Resource Description Framework data model structures data into triples consisting of a subject, predicate, and an object. The subject denotes a thing or concept, typically a URI. An object can be a literal or a URI, while the predicate explains the relationship the subject has with the object. Note, that the predicate need not be reflexive.

RDF data is stored across the Internet as different datasets, each dataset unique to a specific domain of knowledge. These datasets can be categorized into distributed datasets or single point-of-access databases \cite{hausenblas2008size}. The first of these, distributed datasets, is often accompanied with a federated query mediator. This mediator is the entry point for incoming queries. This mediator will then go and fetch the requested information stored on various websites, databases, and namespaces. Since the data is stored by the content creators, no additional data infrastructure is required to store RDF data. However, due to the dependent nature of a distributed dataset, if a single database (hosted by a third party) is slow or offline, the query execution time will be negatively affected. An example of a distributed dataset is the FOAF project \cite{foaf_project}.

Linked data and the Semantic Web gives both humans and machines additional search power. By consolidating duplicate information stored in different databases into a single point-of-access, or centralized cache, we can enable users to perform faster searches with less external dependencies. Since all information is stored on a local (single) system, latency can be reduced and queries can execute deterministically. This thesis presents a centralized RDF database system.

\subsection{MapReduce}

MapReduce \cite{dean2008mapreduce} is a ``programming model and an associated implementation for processing and generating large datasets." It was developed by Jeffrey Dean and Sanjay Ghemawat at Google as an attempt to streamline large-scale computation processes. Internet data can be represented in various data formats, both conventional and unconventional. At Google, this amount of data became so large that processing this data would require ``hundreds or thousands of machines in order to finish in a reasonable amount of time" \cite{dean2008mapreduce}. As a result, MapReduce was created to provide an abstraction of distributed computing and enable programmers to write parallel programs without worrying about the implementation details such as fault tolerance, load balancing, and node communication. The most widely used implementation of the MapReduce framework is Apache Hadoop, described in Section \ref{sec:background_hadoop}.

\begin{figure}[t]
	\centering
	\caption[Overview of MapReduce Execution]
	{Overview of MapReduce Execution \cite{dean2008mapreduce}}
	\label{fig:MapReduce_Execution}
	\includegraphics[width=.8\textwidth]{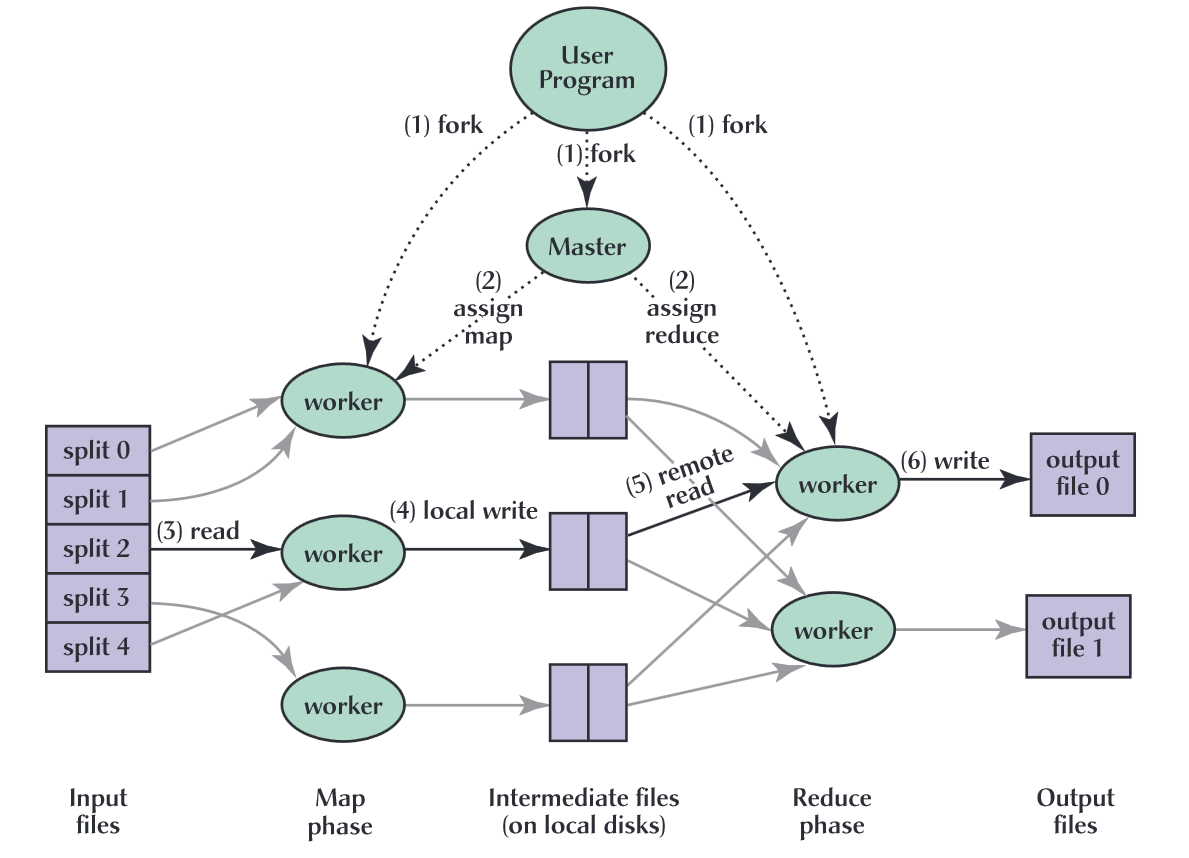}
\end{figure}

The program being executed is divided into many small tasks which are subsequently executed on the various worker nodes. If at any point a sub-task fails, the task is restarted on the same node or on a different node. In the cluster, a single master node exists while the remaining nodes are workers. A worker node can be further classified into mapper nodes and reducer nodes. Mapper nodes perform the map function while reducer nodes perform the reduce task. We will discuss this concept in more detail shortly.

A user program must contain two fundamental components: the \textbf{\emph{map}} function and the \textbf{\emph{reduce}} function. Each of these functions operates on key-value pairs. For the map phase, the key-value pairs are generated from the input file. This is shown as Step 3 in Figure \ref{fig:MapReduce_Execution}. For example, the key could correspond to an ID while the value is a first name. This would give us a key-value pair of $\langle$ID, First Name$\rangle$, e.g. $\langle$001, John$\rangle$, where $\langle$k,v$\rangle$ \, denotes a key-value pair. The keys nor the values need be unique. These key-value pairs are then given to the map function which is described below.

\subsubsection{Map Step}
In the map step, the master node takes the input key-value pairs, divides it into smaller sub-problems, and distributes the task to the worker nodes as shown in Step 2 of Figure \ref{fig:MapReduce_Execution} \cite{apache_hadoop_wiki}. Key-value pairs are assigned to worker nodes depending on several factors such as overall input size, data locality, and the total number of mapper nodes. In the event where the input file is very large, more specifically, where the number of input splits is greater than the number of mapper nodes, then the map jobs will be queued and the mapper nodes will execute the queued jobs sequentially.

The worker node then processes the data and emits an intermediate key-value pair (Steps 3 and 4 of Figure \ref{fig:MapReduce_Execution}). This intermediate key value pair can either be sent a reducer, if a reduce function is specified, or to the master (and subsequently the output file), if there is no reduce function.

\vspace{2 mm}
\alglanguage{pseudocode}
\begin{algorithm}[h]
	\caption{MapReduce Word Count: Map Function}
	\label{algorithm:map_function_example}
	\begin{algorithmic}[1]
		\Procedure{$\mathbf{map}$}{String $input\_key$, String $input\_value$}
		\LeftComment{input\_key: document/file name}
		\LeftComment{input\_value: document/file contents}
		\ForAll{word $w$ in $input\_value$}
		\State EmitIntermediate($w$,``1");
		\EndFor
		\EndProcedure
		\Statex
	\end{algorithmic}
	\vspace{-0.4cm}%
\end{algorithm}

Algorithm \ref{algorithm:map_function_example} shows an example map function of a MapReduce program. This particular example counts the number of times each word occurs in a set of files, documents, or sentences. The input key-value pair is $\langle$document name, string of words$\rangle$. Each mapper node will parse the list of words and emit a key-value pair for every word. The key-value pair is $\langle$word, 1$\rangle$. Note that multiple occurrences of the same word will have the same key-value pair emitted more than once.

\subsubsection{Partition and Shuffle Stages}

MapReduce guarantees that the input to every reducer is sorted by key \cite{oreilly_hadoop}. The process that sorts and transfers the intermediate key-value pairs from mapper to reducers is called the shuffle stage. The method which selects the appropriate reducer for each key-value pair is determined by the partition function.

The partition function should give an approximate uniform distribution of data across the reducer nodes for load-balancing purposes, otherwise the MapReduce job can wait excessively long for slow reducers. A reducer can be slow due to a disproportionate amount of data sent to it, or simply because of slow hardware. The default partition function $\mathcal{F}$ for both MapReduce and Hadoop is:
\begin{center}
	$\mathcal{F}(k,v): hash(k)$ \textbf{mod} $R$
\end{center}
where $R$ is the number of reducers and $k$ is the intermediate key \cite{dean2008mapreduce, apache_hadoop_wiki}. Once the reducer has been selected, the intermediate key-value pair will be transmitted over the network to the target reducer. The user can also specify a custom partition function. Although not shown in Figure \ref{fig:MapReduce_Execution}, the partition function sorts intermediate key-value pairs in memory, on each mapper node, before being written to the local disk. The next step is the shuffle phase.



The shuffle phase consists of each reducer fetching the intermediate key-value pairs from each of the mappers and locally sorting the intermediate data. Each reducer asks the master node for the location of the reducer's corresponding intermediate key-value pairs. Data is then sent across the cluster in a criss-cross fashion, hence the name shuffle. As the data arrives to the reducer nodes, it is loaded into memory and/or disk.

Reducers will continue to collect mapper output until the reducers receive a notification from the JobTracker daemon. The JobTracker process, described in Section \ref{sec:background_hadoop}, oversees and maintains between map outputs and their intended reducer. This process tells the reducer which mapper to ping in order to receive the intermediate data.

Once all the intermediate key-value pairs have arrived at the reducers, the sort stage (within the shuffle phase) begins. Note, a more correct term for this phase would be the \textit{merge} phase as the intermediate data has already been sorted by the mappers. After the intermediate data is sorted locally at each reducer, the reduce step begins. Each reducer can begin this step once it receives its entire task allocation; it need not wait for other reducers.

\subsubsection{Reduce Step}

Each reducer now contains a list of key-value pairs in the form $\langle$key, $\langle$collection of values$\rangle\rangle$. Continuing with our word count example, pseudocode of the reduce function is shown in Algorithm \ref{algorithm:reduce_function_example}. The reducer will execute this function on each key-value pair. If no reducer function is specified in a MapReduce program, then the default reduce implementation is the identity function.

\alglanguage{pseudocode}
\begin{algorithm}[h]
	\small
	\caption{MapReduce Word Count: Reduce Function}
	\label{algorithm:reduce_function_example}
	\begin{algorithmic}[1]
		\Procedure{$\mathbf{reduce}$}{String $key$, Iterator $values$}
		\LeftComment{key: a word}
		\LeftComment{value: a list of counts}
		\State int $sum$ = 0;
		\ForAll{$v$ in values}
		\State $sum$ += ParseInt($v$);
		\EndFor
		\State Emit($key$, AsString($sum$));
		\EndProcedure
		\Statex
	\end{algorithmic}
	\vspace{-0.4cm}%
\end{algorithm}

Inside the function in Algorithm \ref{algorithm:reduce_function_example}, we are summing the values in the $values$ which contains a list of integers represented as strings. The summation of this list is then emitted as the final key-value output pair where they key is the word and the value is the number of times the word occurred all input files.

\subsection{Hadoop} \label{sec:background_hadoop}

Hadoop\footnote{\url{http://hadoop.apache.org}} is an open-source, Java-based implementation of the MapReduce framework. It is developed and maintained by the Apache Software Foundation and was created in 2005. Since Hadoop implements the MapReduce framework, we will describe Hadoop's two components: the MapReduce engine and the Hadoop Distribtued File System (HDFS)\footnote{The explanation in this section and the remainder of this thesis refers to the Hadoop 1.x release.}.

Before we describe the two components, we will briefly describe the possible roles a slave node may be assigned:

\begin{enumerate}
	\item JobTracker - assigns tasks to specific cluster nodes
	\item TaskTracker - accepts tasks from JobTracker, manages task progress
	\item NameNode - maintains file system, directory , and block locations
	\item DataNode - stores HDFS data and connects to a NameNode
\end{enumerate}

In small Hadoop clusters, it is common for the master node to perform all four roles while the slave nodes act as either DataNodes or TaskTrackers. In large clusters, the DataNode and TaskTracker typically assigned to a dedicated node to increase cluster throughput.


\subsubsection{MapReduce Engine}

Hadoop's MapReduce engine is responsible for executing a user MapReduce program. This includes assigning sub-tasks and monitoring their progress. A task is defined as a map, reduce, or shuffle operation.

The JobTracker is the Hadoop service which assigns specific MapReduce tasks to specific nodes in the cluster \cite{apache_hadoop_wiki}. The JobTracker tries to assign map and reduce tasks to nodes such that the amount of data transfer is minimized. Various scheduling algorithms can be used for task assignment. A TaskTracker is a node in the cluster which accepts tasks from the JobTracker. The TaskTracker will spawn multiple sub-processes for each task. If any individual task fails, the TaskTracker will restart the task. Once a task completes, it begins the next task in the queue or will send a success message to the JobTracker.

\subsubsection{Hadoop Distributed File System}
The Hadoop Distributed File System (HDFS) is the distributed file system designed to run on hundreds to thousands of commodity hardware machines \cite{borthakur2007hadoop}. It is highly fault tolerant and was modeled after the Google File System \cite{Ghemawat:2003:GFS:945445.945450}.

HDFS was developed with several goals in mind \cite{apache_HDFS}. Since a Hadoop cluster can scale to thousands of machines, hardware failures are much more common than in traditional computing environments. HDFS aims to detect node failures quickly and automatically recover from any faults or errors. Because of the large cluster size, HDFS is well-suited for large datasets and can support files with sizes of several gigabytes to terabytes.

HDFS implements a master/slave architecture. The HDFS layer consists of a single master, called the NameNode, and many DataNodes, usually one per machine \cite{apache_HDFS}. Each file is split into blocks, scattered across the cluster, and stored in DataNodes. DataNodes are responsible for performing the actual read and write requests as instructed by the NameNode. DataNodes also manage HDFS blocks by creating, deleting, and replicating blocks.

The NameNode is the head of the HDFS system. The NameNode stores all HDFS metadata and is responsible for maintaining the file system directory tree and mapping between a file and its blocks \cite{apache_hadoop_wiki}. It also performs global file system operations such as opening, closing, and renaming files and directories.
In the event the NameNode goes offline, the entire HDFS cluster will fail. It is possible to have a SecondaryNameNode, however this provides intermittent snapshots of the NameNode. Additionally, automatic failover is not supported and requires manual intervention in the event of a NamNode failure.

Data replication is an important feature of HDFS. Just as with standard disks, an HDFS file consists of a sequence of blocks (default size of 64 MB) and are replicated throughout the cluster. The default replication scheme provides triple data redundancy but can be modified as needed. The specific placement of these blocks depends on a multitude of factors but the most significant is a node's network rack placement.

\subsection{HBase}

HBase\footnote{\url{http://hbase.apache.org/}} is an open source, non-relational, distributed database modeled after Google BigTable \cite{chang2008bigtable}. It is written in Java and developed by the Apache Software Foundation for use with Hadoop. The goal HBase is to maintain very large tables consisting of billions of rows and millions of columns \cite{apache_hbase_wiki}. Out of the box, HBase is configured to integrate with Hadoop and MapReduce.

\subsubsection{Data Model}

Logically, the fundamental HBase unit is a column \cite{sever_thesis}. Several columns constitute a row which is uniquely identified by its row key. We refer to a ``cell" as the intersection of a table row and column. Each cell in the table contains a timestamp attribute. This provides us with an additional dimension for representing data in the table.

Columns are grouped into column families which are generally stored together on disk \cite{apache_hbase_wiki}. The purpose of column families is to group columns with similar access patterns to reduce network and disk I/O. For example, a person's demographic information may be stored in a different column family than their biological and medical data. When analyzing biological data, demographic information may not be required. It is important to note that column families are fixed and must be specified during table creation. To include the column family when identifying table cells, we use the tuple: $value =  \langle row\_key, family:column, timestamp \rangle$.


%

\subsubsection{Architecture}

Like Hadoop, HBase implements a master/slave architecture. Tables are partitioned into regions defined by a starting and ending row key. Each of these regions are assigned to an HBase RegionServer (slave). The HBase master is responsible for monitoring all RegionServers and the master is the interface for any metadata changes \cite{apache_hbase_wiki}.

\begin{figure}[h]
	\centering
	\caption[HBase Translation from Logical to Physical Model]{HBase Translation from Logical to Physical Model \cite{oreilly_hbase}}
	\label{fig:hbase_logical_physical}
	\includegraphics[width=0.8\textwidth]{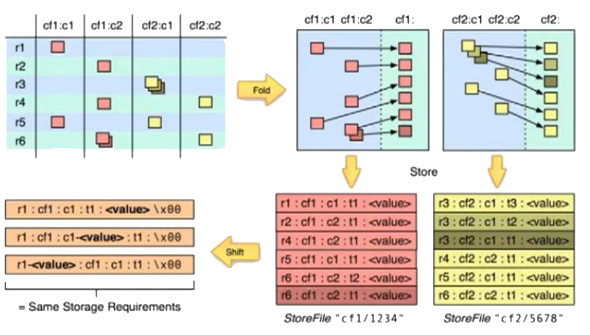}
\end{figure}

In Figure \ref{fig:hbase_logical_physical}, we show the translation from the logical data model to physical representation on disk. In the top left of the figure, an HBase table is shown in its logical form. Stacked squares coming out of the page represent multiple values with different timestamps at a cell. Each column family in the table is split into separated HDFS directories called Stores (top right of Figure \ref{fig:hbase_logical_physical}). Each column family is split into multiple files on disk called StoreFiles. As shown in the bottom right of the figure, each row in the file represents a single cell value stored in the HBase logical model. Each row in the StoreFile contains a key-value pair:
$$\langle key, value \rangle = \langle (row\_key, family:column, timestamp), value \rangle$$
Parenthesis have been added for clarity. Depending on the implementation, the components of the StoreFile key can be swapped without losing information. It is important to note that swapping the components of the compound key allows for more complex database implementations without requiring additional disk space.

\subsubsection{Selected Features} \label{sec:hbase_features}

In this section, we describe some of the advanced features included in HBase, most of which we use in our system. The features we primarily use are HBase's bulk loading tools, bloom filters, and null compression.

Both Hadoop and HBase use the Hadoop Distributed File System. HBase's implementation and coordination of internal services is transparent to Hadoop. As a result, HBase fault tolerance is automatically handled by Hadoop's data replication factor while any RegionServer failures are handled by the HBase cluster services. This simplifies both the setup and management of a Hadoop/HBase cluster.

A very useful feature of HBase is its bulk loading capabilities. As a quick overview: a MapReduce job transforms some input data into HBase's internal file format. We take the output file format from this MapReduce job and load it directly into a live HBase cluster. For specific implementation details and intermediate file formats, we refer the reader to the HBase bulk loading documentation \cite{apache_hbase_bulk_load}.

Bloom filters allow us to reduce the number of HBase rows read during a table scan. HBase offers two types of bloom filters: by row and by row+column. When an HFile is opened, the bloom filter is loaded into memory and used to determine if a given HBase row key exists in the HFile before scanning the HFile. As a result, we are able to skip HBase blocks that do not have the requested values.

Depending on the data model used to store RDF, there may be many empty cells with no values. Because of HBase's physical model, NULL values use zero disk space -- the key-value for a NULL is simply not written to disk. This differs from relational databases where NULL values require space for fixed width fields (e.g. \texttt{char(10)}).

\subsection{Hive}

Developed by Facebook, Hive\footnote{\url{http://hive.apache.org/}} is a data warehouse and infrastructure built on top of Hadoop and provides SQL-like declarative language for querying data \cite{thusoo2009hive}. It is now developed by the Apache Software Foundation and supports integration with HDFS and HBase.

\subsubsection{Data Model} \label{sec:hive_data_model}

Data in Hive is organized into tables, each of which is given a separate directory on HDFS. However, when used in conjunction with HBase, Hive does not duplicate the data and instead relies on the data stored on HDFS by HBase. Assuming we have an existing HBase table, a Hive table is created to manage the HBase table. Creation of a Hive table requires the user to provide a mapping from HBase columns to Hive columns (line 6 of Codebox \ref{code:hive_hbase_ddl}). This is analogous to creating a SQL view over a table where the view is stored in Hive while the table is stored in HBase.

\begin{figure}[h]
	\singlespacing
	\begin{mdframed}
		\begin{lstlisting}[
		label=code:hive_hbase_ddl,
		caption=Hive Table Creation with HBase \cite{hive_hbase_integration},
		]
		CREATE TABLE
		hive_table_1(key int, value string)
		STORED BY
		`org.apache.hadoop.hive.hbase.HBaseStorageHandler'
		WITH SERDEPROPERTIES
		("hbase.columns.mapping" = ":key,cf1:val")
		TBLPROPERTIES
		("hbase.table.name" = "xyz");
		\end{lstlisting}
	\end{mdframed}
\end{figure}

Since HBase stores sequences of bytes as table values, Hive is unaware of the data types in the HBase columns. As a consequence, we are required to specify a primitive type for each Hive column mapping to HBase (line 6 of Codebox \ref{code:hive_hbase_ddl}). Caution must be taken as it is possible to declare a column of the incorrect type. Our Hive table being created in Codebox \ref{code:hive_hbase_ddl} specifies Hive a table referencing two columns of the HBase table \textit{xyz}. One column is of type int and points \textit{xyz}'s row keys while the second column points to the column (and column family) \textit{cf1:val} represented as strings. Line 3 of Codebox \ref{code:hive_hbase_ddl} indicates the table we are creating is stored externally by HBase and will not require storage by Hive. However, Hive will store metadata such as the Hive-HBase column mapping. The DDL query in Codebox \ref{code:hive_hbase_ddl} is executed on Hive to create the table.

\subsubsection{Query Language and Compilation}

Hive provides a SQL-like language called HiveQL as a query language. It is described as SQL-like because contains some of the features such as equi-joins, but it does not fully comply with the SQL-92 standard \cite{oreilly_hadoop}.

The concept of a join in a MapReduce environment is the same as the join operator in a RDBMS. However, the physical plan of a join operator does differ in a MapReduce environment. There are two broad classes of joins in MapReduce: (i) map-side join and (ii) reduce-side join. A reduce-side join is a join algorithm in which the tuples from both relations are compared on their join key and combined during the reduce phase. A map-side join performs the comparison and combination during the map phase.

Hive's default join algorithm is a reduce-side join but it also supports map-side and symmetric hash joins \cite{Wu_query_optimization}. Recent releases of Hive have introduced additional joins such as the Auto MapJoin, Bucket MapJoin, and Skew Join \cite{tang_hive_join_strategies}. However, most of these new algorithms remain map-side joins.

Given an input DDL or DML statement, the Hive compiler converts the string into a query plan consisting of metadata and/or HBase and HDFS operations. For insert statements and queries, a tree of logical operators is generated by parsing the input query. This tree is then passed through the Hive optimizer. The optimized tree is then converted into a physical plan as a directed acyclic graph of MapReduce jobs \cite{thusoo2009hive}. The MapReduce jobs are then executed in order while pipelining intermediate data to the next MapReduce job. An example Hive physical plan can be found in Appendix \ref{sec:appendix_physical_plan1}.

\section{Data Layer} \label{sec:data_layer}

Before we design the schema, we must first understand the characteristics of RDF and SPARQL queries so that we know what optimizations are necessary. In Section \ref{sec:access_patterns}, we present a brief analysis of SPARQL queries and characteristics of common joins. We then describe both our Hive and HBase data model in Section \ref{sec:data_model}. In Section \ref{sec:preproc_data_load}, we outline how data is manipulated and loaded into the distributed file system. Finally, we discuss how we employ bloom filters to take advantage of our schema design in Section \ref{sec:bloom_filters}.

\subsection{Access Patterns}\label{sec:access_patterns}

SPARQL queries access RDF data in many ways. Before we compare and contrast the different data models, we must first understand the primary use cases of our database. Listed below are figures from an analysis of 3 million real-world SPARQL queries on the DBpedia\footnote{\url{http://www.dbpedia.org}} and Semantic Web Dog Food\footnote{\url{http://data.semanticweb.org}} (SWDF) datasets \cite{gallego2011empirical}. We discuss the results of the study and explain key takeaways which we incorporate into our schema and data model.

\begin{figure}[!h]
	\caption[Frequency of Triple Patterns]{Frequency of Triple Patterns (C: Constant, V: Variable) \cite{gallego2011empirical}}
	\centering
	\begin{tabular}{|c|r|r|}
		\hline
		\textbf{Pattern} & \textbf{DBpedia} & \textbf{SWDF} \\ \hline
		C C V & 66.35\% & 47.79\% \\ \hline
		C V V & 21.56\% & 0.52\% \\ \hline
		V C C & 7.00\% & 46.08\% \\ \hline
		V C V & 3.45\% & 4.21\% \\ \hline
		C C C & 1.01\% & 0.001\% \\ \hline
		V V C & 0.37\% & 0.19\% \\ \hline
		C V C & 0.20\% & 0.006\% \\ \hline
		V V V & 0.04\% & 1.18\% \\ \hline
	\end{tabular}
	\label{fig:analysis_triple_patterns}
\end{figure}

We start with the basic analysis of the triple patterns. As shown in Figure \ref{fig:analysis_triple_patterns}, a large percentage of queries have a constant subject in the triple pattern. This indicates that users are frequently searching for data about a specific subject and by providing fast access to an HBase row, assuming the row keys are RDF subjects, we will be able to efficiently serve these queries. Queries with variable predicates constitute about 22\% and 2\% of DBpedia and SWDF queries, respectively. This is more than the percentage of queries with variable subjects. Although variables may appear in any of the three RDF elements (subject, predicate, object), we must take care to balance our data model optimizations for each RDF element.

\begin{figure}[!t]
	\singlespacing
	\centering
	\caption[Number of Joins and Join Type per Query]{Number of Joins and Join Type per Query \cite{gallego2011empirical}}
	\begin{minipage}[c]{0.480\textwidth}
		\includegraphics[width=\textwidth]{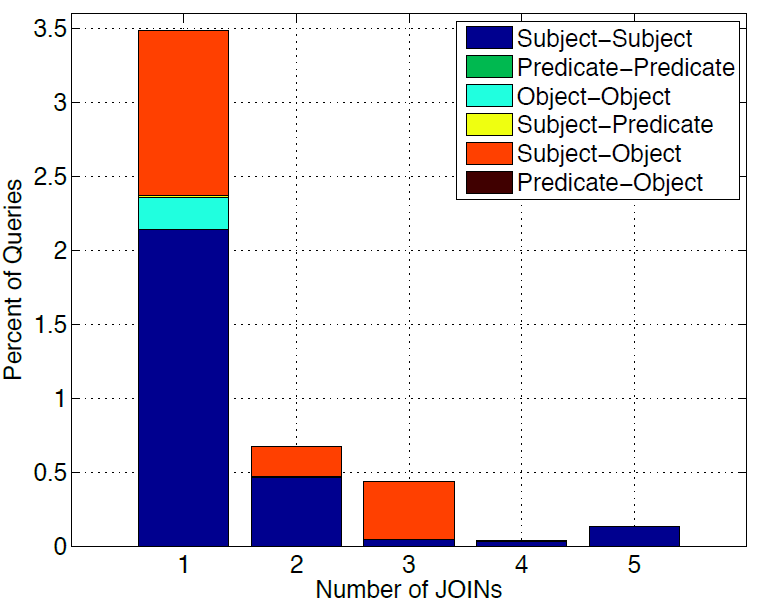}
	\end{minipage}
	\begin{minipage}[c]{0.51\textwidth}
		\includegraphics[width=\textwidth]{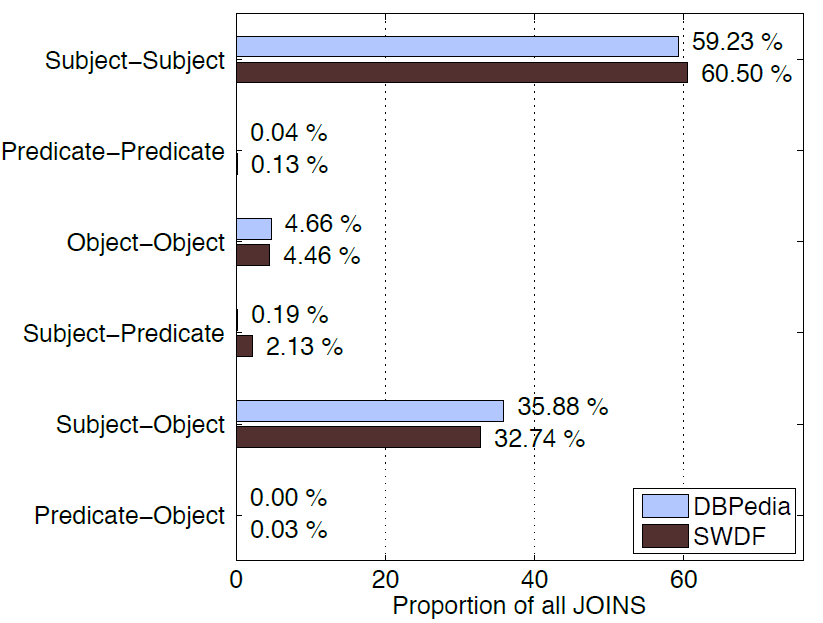}
	\end{minipage}
	\label{fig:analysis_joins}
\end{figure}

Moving onto an analysis of the queries with joins, in the left graph of Figure \ref{fig:analysis_joins}, we see that 4.25\% of queries have at least one join. We define a join as two triple patterns which share a variable in any of the three RDF positions. Subject-subject joins constitute more than half of all joins while subject-object constitute about 35\%. Combined with the results from the constant/variable patterns, we can conclude that subjects are by far the most frequently accessed RDF element. Therefore, we hypothesize that optimizing the data model for fast access of RDF subjects will allow us to efficiently answer queries requesting subjects and joins involving subjects.

\subsection{Schema Design}\label{sec:data_model}

There are several data models for representing RDF data which have been studied and implemented \cite{hexastore, wilkinson2006jena, abadi2007scalable}. Taking into account the analysis performed in the previous section, we employ a property table as our data model. In this section we will explain our rationale and compare our schema with other models.

A property table derives its name from the logical representation of the table. Each row contains a key and each column stores some property of that key.
Using HBase as our storage mechanism and Hive as our query engine, we employ a property table with the following characteristics:

\begin{enumerate}
	\item Subjects are used as the row key.
	\item Predicates are represented as columns.
	\item Objects are stored in each cell.
	\item Each cell can store multiple values, distinguished by their timestamp.
	\item The property table is non-clustered. We use a single property table opposed to having multiple property tables grouped by similarity.
\end{enumerate}

\noindent Our rationale for selecting a property table stems from the shortcomings of traditional triple-tables. A triple-table is a single table with three columns: a subject column, predicate column, and object column. Consider the SPARQL query in Codebox \ref{self-join_code}. In a triple-table, we must find each subject that satisfied the three graph patterns specified in the SPARQL query. This requires us to join each subject/row with itself. A single self-join will allow us to identify graph patterns with two triples. For the query in Codebox \ref{self-join_code}, we need two self-joins. Since we must check for all subjects that match the triple patterns in the query, we must join the entire triple table with itself. Generalizing this case: for a SPARQL query with $n$ triple patterns, we must perform $n-1$ self-joins.

\begin{figure}[t]
	\caption{SPARQL Query and Graph Representation with Self-Joins}
	\vspace{-4 mm}
	\begin{minipage}{0.6\textwidth}
		\singlespacing
		\begin{mdframed}
			\begin{lstlisting}[label=self-join_code,caption=SPARQL Query]
			SELECT ?firstName
			WHERE {
			?x foaf:firstName ?firstName .
			?x foaf:age 30 .
			?x foaf:livesIn "Austin"
			}
			\end{lstlisting}
		\end{mdframed}
	\end{minipage}
	\begin{minipage}{0.39\textwidth}
		\includegraphics[width=0.9\textwidth]{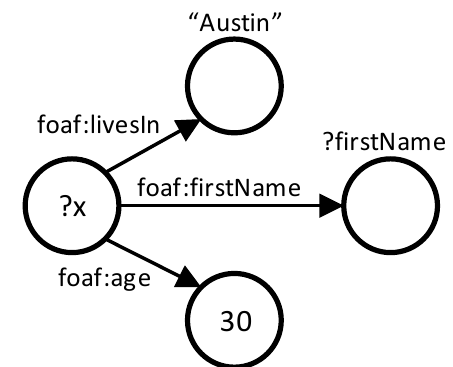}
	\end{minipage}
	\label{fig:sparql_query_self_joins}
\end{figure}

For queries with a large number of self-joins, a triple-table approach may become infeasible. A property table eliminates the self-join requirement. In a property table, all object data is stored in the same row as its subject -- with the predicate dictating the column placement. As a result, we can retrieve all objects associated with a subject in a single row access. The SPARQL query in Codebox \ref{self-join_code} will require zero self-joins in a property table approach. The elimination of self joins is a major advantage of our approach.

Another approach for storing RDF is to vertically partition the data \cite{abadi2007scalable}. This requires creating $n$ tables where $n$ is the number of unique properties, or predicates, and each table contains two columns: the subject and the object. It has been discovered that a major disadvantage of property tables is the overhead due to NULL values \cite{abadi2007scalable}. However, since HBase does not store NULLs on disk, our system does not have this liability. The strength of property tables lie in fast access times for multiple properties regarding the same subject. In a vertically partitioned datastore, the SPARQL query in Codebox \ref{self-join_code} would require accessing three different tables and require joins.

The Hexastore approach for storing RDF builds six indices, one for each of the possible ordering of the three RDF elements \cite{hexastore}. Queries are executed in the microsecond range and outperform other RDF systems by up to five orders of magnitude. Although Hexastore has stellar performance, it requires indices to be constructed and stored in memory. As the dataset size increases, memory usage increases linearly. As a result, memory usage becomes the limiting factor of Hexastore's ability to accommodate large datasets. For this reason, we do not adopt the Hexastore approach.
\\

\subsection{Preprocessing and Data Loading} \label{sec:preproc_data_load}

Now that we have selected our data model, we will talk about how data is loaded into the system. Data loading consists of two phases: the creation of the HBase table and the loading of data into the cluster.

\begin{figure}[t]
	\centering
	\begin{minipage}{0.39\textwidth}
		\caption{HBase Key Generation}
		\includegraphics[width=\textwidth]{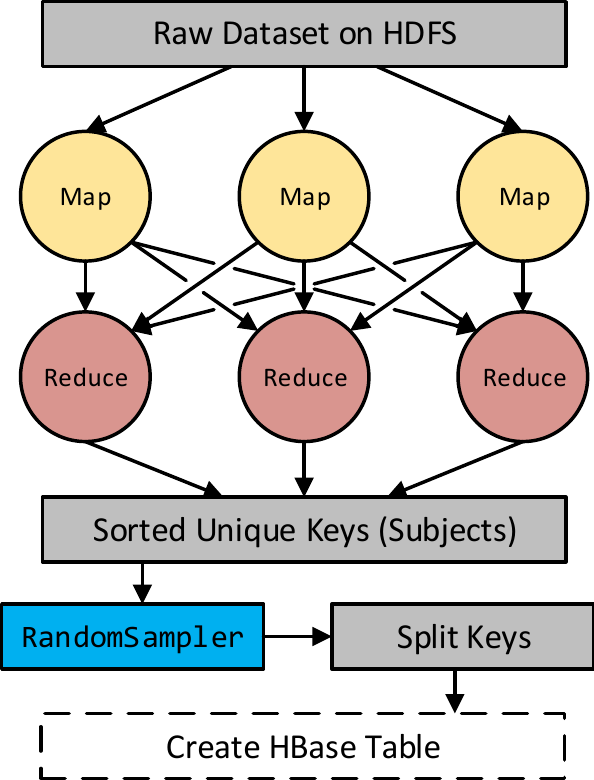}
		\label{fig:key_generation}
		\vspace{3.5 mm}
	\end{minipage}
	\begin{minipage}{0.05\textwidth}
		\hspace{1.5 mm}
	\end{minipage}
	\begin{minipage}{0.46\textwidth}
		\caption{HFile Creation and Upload}
		\includegraphics[width=\textwidth]{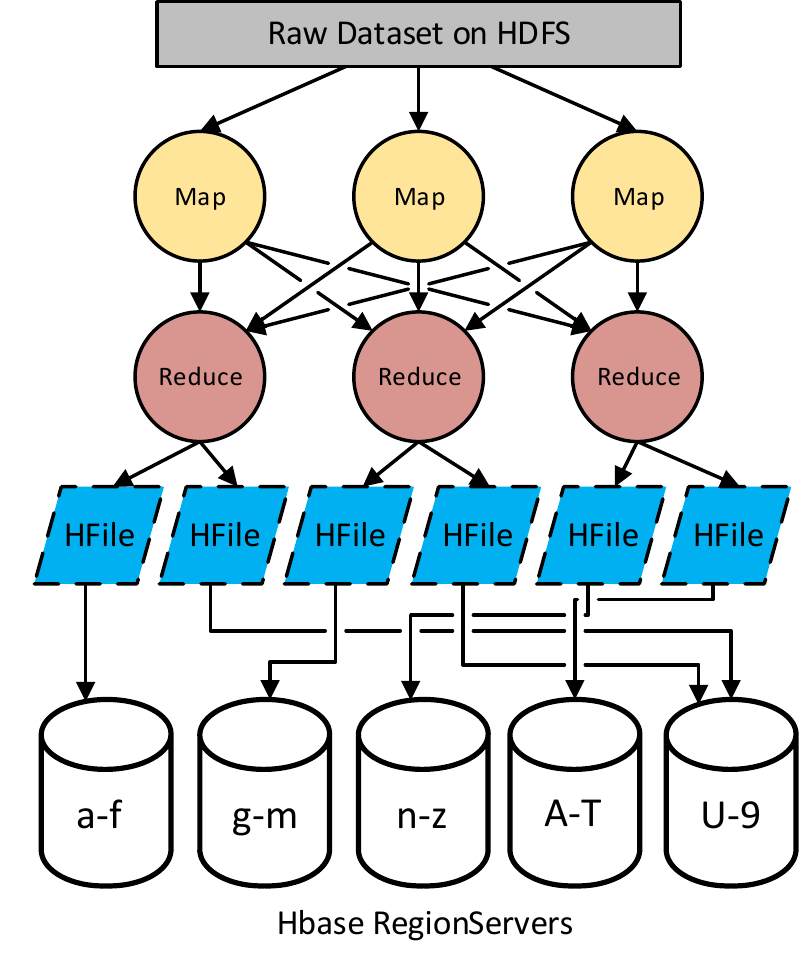}
		\label{fig:hfile_create_load}
	\end{minipage}
\end{figure}

When we bulk load into HBase, keys are sent to different slave nodes depending on a set of \textit{split-keys}. As a result before we can create the HBase table, we must identify these split keys (unique subject keys) which results in the dataset being relatively balanced when stored across the cluster. Due to the nature of our bulk loading technique, this step must be done first. There are several approaches to compiling a list of unique keys. We implement a MapReduce program which extracts all keys in the map phase and eliminates duplicates in the reduce phase. Once we have the unique keys, we use Hadoop's \texttt{InputSampler.RandomSampler} to read and draw samples from the list of unique keys. This process is illustrated in Figure \ref{fig:key_generation}. For small datasets it may be possible to do this sequentially on a single machine, however, for large datasets, it is recommended to process the input in parallel. We use the following parameters for the \texttt{InputSampler.RandomSampler}:

\begin{enumerate}
	\item The probability with which a key will be chosen is 10\%.
	\item The total number of samples to obtain is 10\% of the number of keys.
	\item The maximum number of sampled input file-splits is 50\%.
\end{enumerate}

\noindent We have found that the above parameters balance accuracy and compute time. The \texttt{InputSampler.RandomSampler} will output a list of sorted keys. We take this list of keys and select 2, 4, 8, or 16 keys that evenly split the list into partitions. These become the \textit{split-keys}. We use powers of two since those are the number of nodes used in our experiment. When creating the HBase table, we give these split-keys as additional parameters. The effect of providing these keys to HBase become apparent during the next phase: loading the data into the table.

Loading data into HBase is a two step process. First, we must create a set of files which match HBase's internal storage format. We do this with a custom MapReduce job. However, other methods are provided by HBase which allow various input formats \cite{apache_hbase_bulk_load}. Inside our MapReduce program, we parse the input file and perform various transformations to the dataset. One of these transformations includes compressing all URIs using their corresponding namespace prefix.

Another critical transformation is converting special RDF data types to primitive types. Datasets may include non-primitive types such as longitude, elevation, and hour. Due to the unstructured representation of RDF data, data type information is included in the RDF object itself (of type string) as shown below:
\begin{verbatim}
"6"^^<http://www.w3.org/2001/XMLSchema#integer>
\end{verbatim}
When we convert from the above string to a primitive type, we must remove the additional information from the string itself and infer its primitive type. It is important to note we are not losing this information as we are including this information as the RDF object's type, which is later used by Hive. The transformed object becomes \texttt{6} of type integer. In the event the Hive or HBase system loses the type information, a mapping of predicates to predicate types is also maintained and written to disk.

The above transformations are done in the map phase which outputs an HBase \texttt{Put}. A \texttt{Put} is HBase's representation of a pending value insertion. Conveniently by the inherent nature of MapReduce, all \texttt{Put}s for a single subject are sent to the same reducer since key-value pairs are sorted on the key (subject). The reduce phase inserts the \texttt{Put}s into an HFile. An HFile is HBase's internal data storage format.

The final phase of bulk loading requires moving the HFiles to the appropriate HBase RegionServers, as shown in Figure \ref{fig:hfile_create_load}. HBase does this by iterating through the output HFiles generated by the MapReduce job located on HDFS. HBase compares the first key in the HFile and determines which RegionServer it belongs to. This mapping between keys and RegionServers is determined by the split-keys generated previously and HBase table metadata (Figure \ref{fig:key_generation}). The entire bulk loading process can be performed on a live Hadoop/HBase cluster. Once the above steps are complete, the data becomes live in HBase and is ready to be accessed.

We have explained the preprocessing required when loading data into HBase. This included two steps: (i) identifying the set of HBase region split-keys and (ii) transforming the input data according to a set of rules and moving the data into the RegionServers.

\subsection{Bloom Filters} \label{sec:bloom_filters}

As mentioned in Section \ref{sec:hbase_features}, HBase allows the use of bloom filters. A bloom filter is a probabilistic data structure used to test if an element is part of a set or not. Usually represented as a bit vector, it returns ``possibly in the set" or ``definitely not in the set" \cite{Bloom:1970:STH:362686.362692}. The bits in the vector are set to 0 or 1 depending on a set of hash functions applied to the element being added/searched in the bloom filter.

Each HFile (or block), represents a fragment of an HBase table. Each block is equipped with a bloom filter. As a result, we are able to quickly check whether a table value exists in block or not. Upon inserting data into HBase, we hash the row key and column as a compound hash key (in RDF terms, the subject and predicate), and populate our bloom filter. When searching for RDF triples in the HBase table, we hash the subject and predicate and check the bloom filter. The bloom filter will tell us whether an RDF object exists in the block with the (hashed) subject and predicate. If the bloom filter tells us no, then we skip reading the current block.

We can achieve greater gains from the use of bloom filters through our bulk loading process with MapReduce. If MapReduce was not used for bulk loading and data was inserted randomly (data is generally sorted randomly in input files), then all triples for a specific subject ``albert" will be scattered throughout many blocks across the HDFS cluster. We define an ``albert" triple an RDF triple with ``albert" as the subject. When scanning the HBase table, we would see many true results from our bloom filters since the ``albert" triples lie in many HDFS blocks. By using MapReduce, data is sorted before the reduce stage and therefore all triples with ``albert" as the subject will be sent to the same reducer and as a result, be written to the same HFiles. This allows our bloom filters to return true for these ``albert"-triple concentrated files and return false for the other HDFS blocks.

Additionally, by using a compound hash key consisting of the row key and column, we can further reduce the number of blocks read. HBase offers two types of bloom filters with different hash keys: (i) row key only as the hash key and (ii) row key concatenated with the column. Continuing from our previous example with ``albert" triples, assume all ``albert" triples lie in two blocks. Consider the case where we are looking for the city ``albert" lives in (denoted by the predicate ``livesInCity"). Note that there is only one possible value for the object because a person can only live in one city at a time. If we employ a row-key-only bloom filter, then the bloom filter will return true for both blocks and we must scan both. However, if we use a row key and column hash, then the bloom filter will return false for one block and true for the other -- further reducing the number of blocks read. This feature is especially useful for predicates with unique value as there is only one triple in the entire database with the value we require. In the ideal case, bloom filters should allow us to read only a single HFile from disk. However, due to hash collisions and false positives, this may not be achieved.

The above techniques allow us to accelerate table scans and by skipping irrelevant HDFS blocks. In the next section, we describe how we use Hive to query data stored in HBase.

\section{Query Layer} \label{sec:query_layer}

We use Apache Hive to execute SPARQL queries over RDF data stored in HBase. Several processes must completed before the query is executed as a MapReduce job. In Section \ref{sec:join_conditions}, we describe how the query is reformulated to leverage the advantages created by our property table design. This requires creating intermediate views described in Section \ref{sec:views}. From there we build the abstract syntax tree in Section \ref{sec:ast_construct}. The final component of the query layer is the HiveQL generator, which we discuss in Section \ref{sec:hql_generation}.

\subsection{Join Conditions}\label{sec:join_conditions}
\begin{figure}[t]
	\singlespacing
	\centering
	\begin{minipage}[c]{0.495\textwidth}
		\caption{Join Query (Graph Form)}
		\centering
		\includegraphics[width=.8\textwidth]{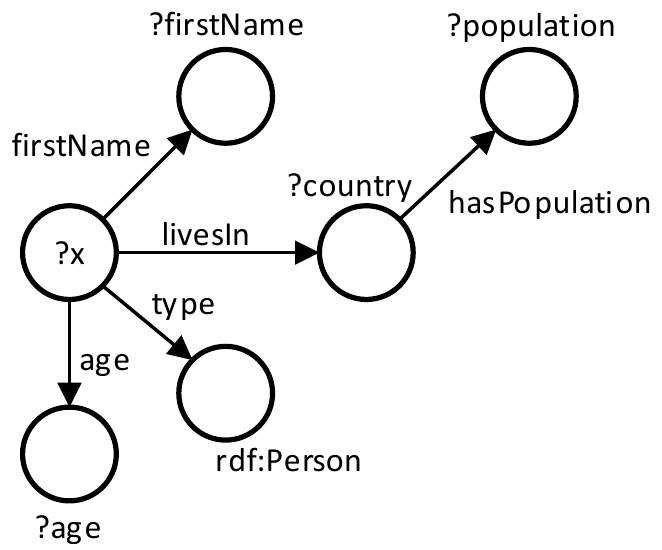}
		\label{fig:join_query_graph}
	\end{minipage}
	\begin{minipage}[c]{0.495\textwidth}
		\caption{Join Query (Join Condition)}
		\centering
		\includegraphics[width=.8\textwidth]{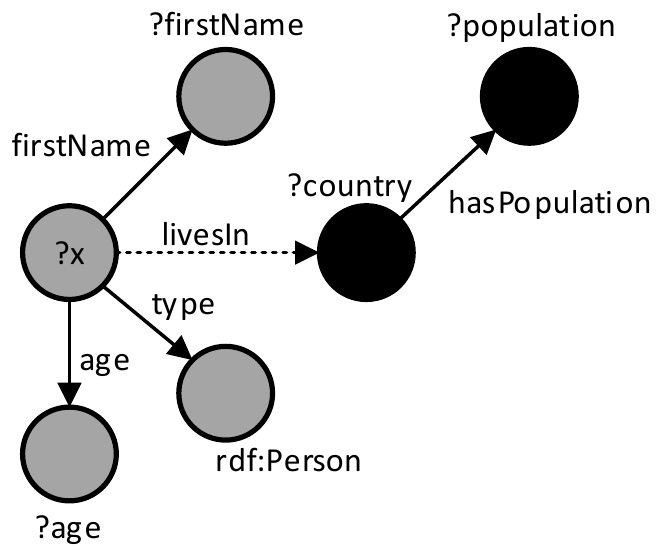}
		\label{fig:join_query_join}
	\end{minipage}
\end{figure}

We define a join as the conjunction of two triple patterns that share a single variable in different RDF positions -- specifically, a variable in the subject position and a variable in the object position. This is different from the analysis performed in Section \ref{sec:access_patterns}. Triple patterns that share the same variable subject do not require a join in our data model due to the nature of property tables. The SPARQL query defines a set of triple patterns in the \texttt{WHERE} clause for which we are performing a graph search over our database and identifying matching subgraphs. In order to utilize the features of the property table, we must first group all triple patterns with the same subject. Each group will become a single Hive/HBase row access. This is done in three steps. We explain the process of identifying a join query using the query in Codebox \ref{code:sparql_join_query}, also depicted as a graph in Figure \ref{fig:join_query_graph}.

\begin{figure}[h]
	\singlespacing
	\begin{mdframed}
		\begin{lstlisting}[label=code:sparql_join_query,caption=Join Query (SPARQL)]
		PREFIX foaf: <http://xmlns.com/foaf/0.1/>
		PREFIX rdf:  <http://www.w3.org/1999/02/22-rdf-syntax-ns#>

		SELECT ?firstName, ?age, ?population
		WHERE {
		?x foaf:firstName ?firstName .
		?x foaf:age ?age .
		?x rdf:type rdf:Person .
		?x foaf:livesIn ?country .
		?country foaf:population ?population
		}
		\end{lstlisting}
	\end{mdframed}
\end{figure}

\begin{figure}[h]
	\centering
	\caption{Mapping of Query Subjects to Triple Patterns}
	\includegraphics[width=.5\textwidth]{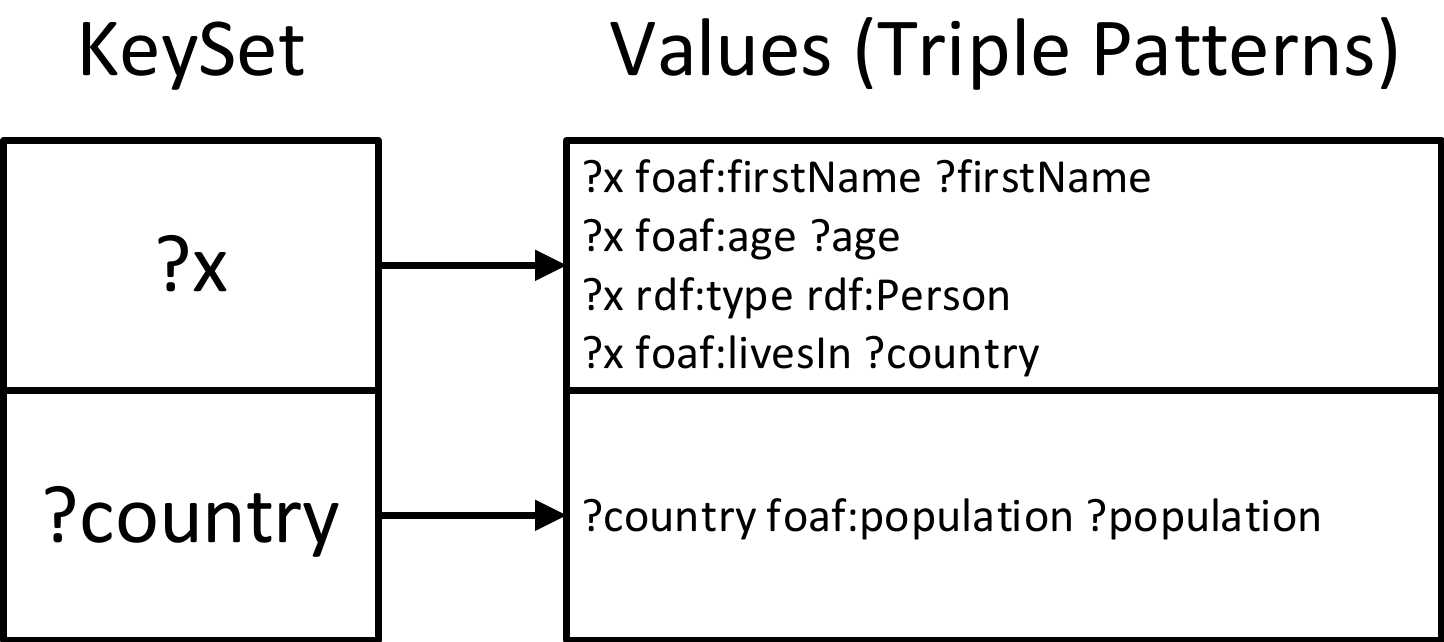}
	\label{fig:join_keyset}
\end{figure}

First, we must group all triple patterns in the SPARQL \texttt{WHERE} clause by subject. The query in Codebox \ref{code:sparql_join_query} conveniently satisfies this property by default. We then iterate through the triple patterns and create a list of unique subjects that are being referenced. In Codebox \ref{code:sparql_join_query}, our list of unique subjects, $L$, contains two elements: $L=\{?x, ?country\}$. In the event we have the same constant subject occurring in two different positions in two triple patterns, we must perform a join. This is demonstrated in Codebox \ref{code:bsbm_q7_sparql}. Using $L$, we create a mapping such that each unique subject $S^* \in L$ maps to a list of triple patterns of the form $\langle S^*,P,O \rangle$. The result is a map $M$ with $KeySet(M)=L$, shown in Figure \ref{fig:join_keyset}.

Second, we must check if any two triple patterns in the \texttt{WHERE} clause have the form $\langle S_1 P_1 O_1 \rangle$ and $\langle S_2 P_2 O_2 \rangle$ where $O_1=S_2$. This is done by iterating through all triple patterns in our mapping. For each triple pattern, we check if the object matches any object in our list of unique subjects, $KeySet(M)$. If a triple pattern, $\langle S,P,O \rangle$, contains an $O \in KeySet(M)$, then we know that a join must be performed on $S$ and $O$. Figure \ref{fig:join_query_join} illustrates the mapping in Figure \ref{fig:join_keyset} in graph form. The different node colorings represent the different objects being referenced by a specific subject. The dashed line represents line 9 of Codebox \ref{code:sparql_join_query} which is the triple pattern indicating a join is necessary. In our property table model, this requires a join. In other data models, a join may not be required and may not be caused by line 9. We summarize the process for identifying a join condition: Let $S$ be a subject in the query $Q$. Let $N$ be the number of triple patterns in the \texttt{WHERE} clause of the query. Let: $M:S\rightarrow \{ \langle S, P_1, O_1 \rangle, ..., \langle S, P_n, O_n \rangle \}, n \leq N$. A SPARQL query will require a join in our system if the SPARQL \texttt{WHERE} clause satisfies the following properties:
\begin{enumerate}
	\singlespacing
	\item $|KeySet(M)| \geq 2$
	\item $\exists \langle S_1, P_1, O_1 \rangle, \langle S_2, P_2, O_2 \rangle \in Q$ such that $\Big(O_1=S_2\Big) \wedge \Big(S_1,S_2 \in KeySet(M)\Big)$
\end{enumerate}

\subsection{Temporary Views}\label{sec:views}

An abstract syntax tree includes relations to which we apply relational operators. Before we can construct the AST, we must define these relations. This is done by creating a \textit{temporary view} by using Hive. We say \textit{temporary} because the term materialized would be inappropriate since Hive is not duplicating or caching the view data. No additional disk space is used (excluding metadata) since we reference the single ``master" HBase table as outlined in Section \ref{sec:hive_data_model}. For the remainder of this thesis, the term view and table are used interchangeably as we refer to Hive tables.

\begin{figure}[t]
	\centering
	\caption{Hive View Generation}
	\includegraphics[width=.99\textwidth]{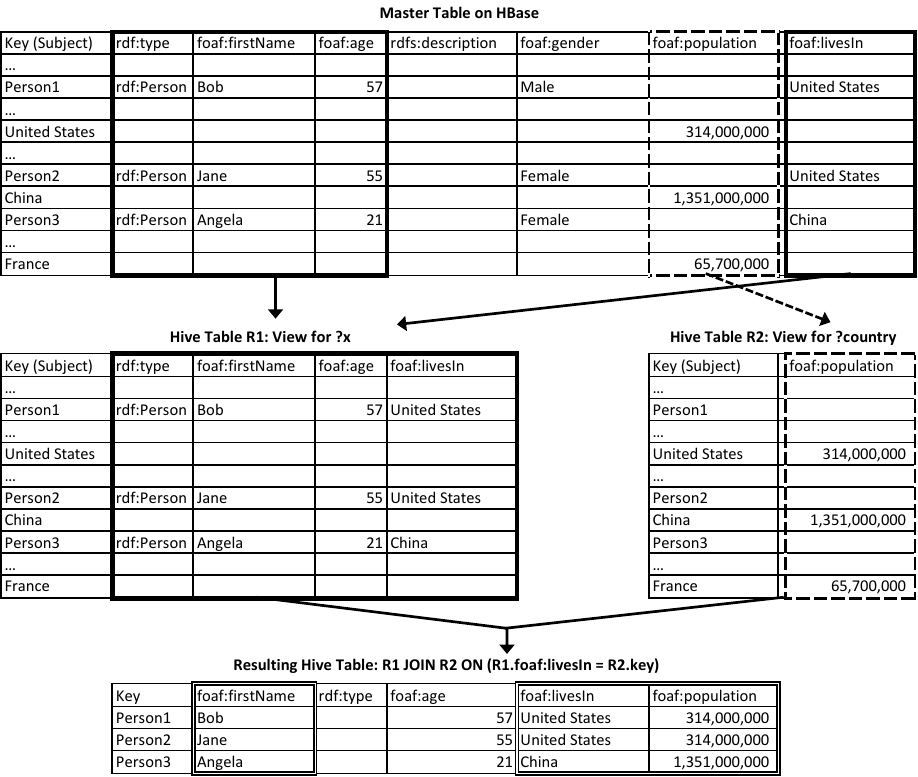}
	\label{fig:hive_views}
\end{figure}

Revisiting our subject-triple mapping in Figure \ref{fig:join_keyset}, we can see the triple patterns associated with each unique subject. Using the same SPARQL query in Codebox \ref{code:sparql_join_query}, we want to create views for the two subjects, \texttt{?x} and \texttt{?country}. Consider \texttt{?x} first. As shown in Figure \ref{fig:join_keyset}, we see that \texttt{?x} has four triple patterns we must match in the database. We can find the object values by accessing the table row and retrieving the value at each of the four predicate columns. Note: we only need to access four columns for information about \texttt{?x}, thus we can omit all other columns because they will not be referenced by any triple pattern about \texttt{?x}. Therefore, we can create a view over the master HBase table for \texttt{?x} with four columns plus one column for the row key. The four columns are specified by the predicate of each of the triple patterns in $M(?x)$.

We can generalize this and make the statement: for each unique subject $S^*$ in a SPARQL query, we can create a view $V$ over the HBase table such that for each column $c \in V$ if $c\neq row\_key$, then $c$ is also a predicate in $M(S^*)$. This is illustrated in Figure \ref{fig:hive_views}. The thick solid line indicates columns mapped by the view for \texttt{?x}. Since the Hive to HBase column mapping is handled when we create the external Hive table, these columns need not be adjacent in HBase. In this view, we see the four predicates in the SPARQL query that have \texttt{?x} as their subject. Since \texttt{?x} is a variable, \texttt{?x} assumes the value of the row key. We apply this same process to the view for \texttt{?country}. We index each subject in $KeySet(M)$ with an integer and create the corresponding Hive table named after the integer index. For example, \texttt{R2} refers to the view for \texttt{?country}. Hive performs the join and we see the joined relation at the bottom of the figure. The columns with a double border are projected as the final query result.

While the SPARQL/Hive query is being executed, the views consume disk space for metadata. Once the query completes, all views are discarded and their metadata is deleted. It is possible for the views to persist but since each view contains a mapping to HBase columns for a unique SPARQL query, it is unlikely that future queries will reference the exact same data.

By nature of SPARQL, given a specific subject and predicate, it is possible for multiple objects to exist. In HBase, these objects are distinguished by their timestamp value. Since Hive does not support cell-level timestamps, we store multivalued attributes in Hive's array data type. The default data type for Hive columns are strings. For numeric types, the system checks the predicate-primitive type mapping (see Section \ref{sec:preproc_data_load}) before creating the Hive view.

The primary benefit of applying the relational concept of views to Hive is the reduction of information processed by Hadoop/MapReduce. By using Hive, we are able to perform a relational project on the underlying HBase table and exclude irrelevant information to the SPARQL query. Views serve as an optimization by reducing data processed by mapper nodes and by reducing data shuffled by the network.

\subsection{Abstract Syntax Tree Construction} \label{sec:ast_construct}

Given the subject-triple pattern mapping $M$ described in the previous section, we are almost ready to construct the abstract syntax tree. The remaining step is to identify the Hive columns associated with the remaining variables. Continuing our example in Codebox \ref{code:sparql_join_query}, we must assign a Hive column to \texttt{?population, ?age,} and \texttt{?firstName}. This is easily done by looking at the predicate for that triple pattern. \texttt{?firstName} is the data stored in the column \texttt{foaf:firstName}, \texttt{?age} is the column \texttt{foaf:age} and so forth. We have now mapped every variable in the SPARQL query to a corresponding Hive column.

Using Apache Jena, we construct an AST using extended relational algebra notation to denote operators: $\bowtie$ is the join operator, $\pi$ is the projection operator, $\sigma$ is the selection operator, $\delta$ is the duplicate elimination operator, and $\tau$ is the sorting operator.

\begin{figure}[t]
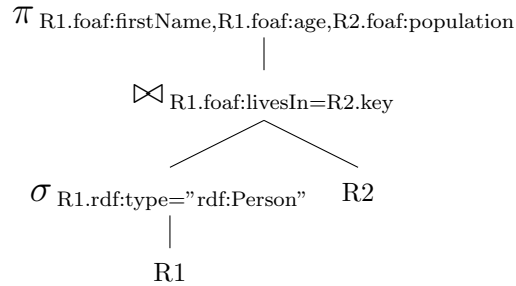

	\caption[AST for Sample SPARQL Query]{AST for SPARQL Query in Codebox \ref{code:sparql_join_query}}
	\Large
	\Tree [.$\pi_{\:\textrm{\scriptsize R1.foaf:firstName,R1.foaf:age,R2.foaf:population}}$ [.$\bowtie_{\:\textrm{\scriptsize R1.foaf:livesIn=R2.key}}$ [.$\sigma_{\:\textrm{\scriptsize R1.rdf:type="rdf:Person"}}$ $\textrm{\small R1}$ !{\qbalance} ] [.$\textrm{\small R2}$ ]]]
	\label{fig:join_query_ast}
\end{figure}

The newly constructed AST is shown in Figure \ref{fig:join_query_ast}. In the top-most projection, we explicitly tell Hive which relation to retrieve the columns from. Although this is not required for this specific query, for complex queries, the added clarity makes the final HiveQL easier to read for debugging and development purposes. For the remainder of this section, we describe how we convert a SPARQL query, including advanced features such as \texttt{OPTIONAL} clauses, into an AST.

\begin{figure}[h]
	\singlespacing
	\begin{mdframed}
		\begin{lstlisting}[label=code:bsbm_q7_sparql,caption=Berlin SPARQL Benchmark Query 7]
		PREFIX rdfs: <http://www.w3.org/2000/01/rdf-schema>
		PREFIX rev: <http://purl.org/stuff/rev>
		PREFIX foaf: <http://xmlns.com/foaf/0.1/>
		PREFIX bsbm: <http://www4.wiwiss.fu-berlin.de/bizer/bsbm/v01/vocabulary
		PREFIX dc: <http://purl.org/dc/elements/1.1/>

		SELECT ?productLabel ?offer ?price ?vendor ?vendorTitle ?review ?revTitle ?reviewer ?revName ?rating1 ?rating2
		WHERE {
		%ProductXYZ% rdfs:label ?productLabel .
		OPTIONAL {
		?offer bsbm:product %ProductXYZ% .
		?offer bsbm:price ?price .
		?offer bsbm:vendor ?vendor .
		?vendor rdfs:label ?vendorTitle .
		?vendor bsbm:country <http://downlode.org/rdf/iso-3166/countries#DE>
		?offer dc:publisher ?vendor .
		?offer bsbm:validTo ?date .
		FILTER (?date > %currentDate% )
		}
		OPTIONAL {
		?review bsbm:reviewFor %ProductXYZ% .
		?review rev:reviewer ?reviewer .
		?reviewer foaf:name ?revName .
		?review dc:title ?revTitle .
		OPTIONAL { ?review bsbm:rating1 ?rating1 . }
		OPTIONAL { ?review bsbm:rating2 ?rating2 . }
		}
		}
		\end{lstlisting}
	\end{mdframed}
\end{figure}

We now construct an AST for Query 7 from the Berlin SPARQL Benchmark \cite{bizer2009berlin}. This is a very complex query involving a 5-way join, filter operations, and nested optional clauses. The query can be found in in Codebox \ref{code:bsbm_q7_sparql} and the full AST is shown in Figure \ref{fig:ast_bsbm_7}. The unique subjects and their associated views are shown in the figure's caption. It is important to note that $R1...R5$ contain a subset of columns from the master HBase table as outlined in Section \ref{sec:views}. Due to space constraints, we omit namespace prefixes and full URIs from the AST. Anything in the form \texttt{"\%xyz\%"} represents a constant.

\begin{figure}[t]
	\caption[AST for Berlin SPARQL Benchmark Query] {
		AST for Berlin SPARQL Benchmark Query 7 \\
		\noindent
		\footnotesize Subject-View Mapping: R1$\leftarrow$?product, R2$\leftarrow$?offer, R3$\leftarrow$?vendor, R4$\leftarrow$?review, R5$\leftarrow$?reviewer
	}
	\Large
	\Tree [.$\pi_{\:\textrm{\tiny productLabel,offer,price,vendor,vendorTitle,review,revTitle,reviewer,revName,rating1,rating2 }}$ [.$\sigma_{\:\textrm{\tiny R2.validTo}\: > \: \textrm{\tiny``\%currentDate\%" }}$ [.$\leftouterjoin_{\:\textrm{\tiny R1.key=R2.product }}$ [.$\sigma_{\:\textrm{\tiny R1.key=``\%ProductXYZ\%"}}$ $_{\textrm{\small R1 }}$ ] [.$\bowtie_{\:\textrm{\tiny R2.vendor=R3.key }}$ [.$_{\textrm{\small R2 }}$ ] [.$\leftouterjoin_{\:\textrm{\tiny R4.key=R1.reviewFor }}$ [.$\sigma_{\:\textrm{\tiny R3.country=``DE" }}$ $_{\textrm{\small R3 }}$ ] [.$\leftouterjoin_{\:\textrm{\tiny R4.reviewer=R5.key }}$ $_{\textrm{\small R4 }}$ $_{\textrm{\small R5 }}$ !{\qbalance} ] ] !{\qframesubtree} ] ] ] ]
	\label{fig:ast_bsbm_7}
\end{figure}

The \texttt{OPTIONAL} clause in SPARQL is a binary operator that combines two graph patterns \cite{sparql_optional}. We define the \textit{basic query graph} as the graph pattern formed by the SPARQL query excluding optionals and the \textit{extended query graph} formed by including the optionals. We define the \textit{database graph} as the graph pattern in the database currently being examined by our query engine.

When evaluating \texttt{OPTIONAL}, we have the following cases:
\paragraph{Case 1} The database graph does not match the basic query graph. In this case, the database graph is not included in the result set.
\paragraph{Case 2} The database graph matches the basic query graph but not the extended query graph. As a result, the database graph is included in the result set.
\paragraph{Case 3} The database graph matches the basic query graph and the extended query graph but fails \texttt{OPTIONAL} conditionals. In this case, the database graph is not included in the result set.
\paragraph{Case 4} The database graph matches the basic query graph and the extended query graph and passes all \texttt{OPTIONAL} conditionals. Only if the database graph satisfies everything, then we return the database graph in the result set.

In the traditional relational model, Case 4 can be accommodated with an outer join. Because most SPARQL queries specify the required graph patterns before the optional patterns, it is common for left outer joins to be used. This is best illustrated by the boxed subtree in Figure \ref{fig:ast_bsbm_7}. Both R4 and R5 contain information about reviews and reviewers. Since no filter is applied to R4 or R5, we return all values. However, when we perform a filter on the R3 (vendors) we must remove non-German (DE) countries. If a vendor has no country listed, we must not remove them from the result set. The selection $\sigma_{\:\textrm{\small R2.validTo}\: > \: \textrm{\small ``\%currentDate\%"}}$ appears at the top of the tree because Hive only supports equality conditionals within a join condition. We traverse the AST and check if any non-equality condition appears below a join -- usually a selection operator. We then move the operator to appear above the joins in the AST. Logically and syntactically, these selections are performed after the joins complete.

Given the AST, we can then perform additional transformations or optimizations such as attribute renaming or selection push-down. Since the AST resembles a traditional relational AST, we can apply the same set of RDBMS optimization rules to our translator. Since this system is in its early stages, we do not include a full-fledged optimizer but instead optimize selection placement in the AST. As shown in Figure \ref{fig:ast_bsbm_7}, all equality selections occur before any joins are performed.

\subsubsection{Additional SPARQL Operators}

The Berlin Query 7 demonstrates use of the \texttt{FILTER} and heavy use of the \texttt{OPTIONAL} keyword. Our parser and AST generator account for other SPARQL operators as well. For the remainder of this section, we outline the process for identifying each operator and its operands, how we transform it into relational algebra, and emit the correct HiveQL code. The additional features we support are: \texttt{ORDER BY}, \texttt{FILTER} (with regular expressions), \texttt{DESCRIBE}, \texttt{LIMIT}, \texttt{UNION}, \texttt{REDUCED}, and \texttt{BOUND}.

\texttt{ORDER BY}. Located near towards the end of each SPARQL query, the \texttt{ORDER BY} clause appears at the top of the AST, denoted by $\tau$. In Figure \ref{fig:ast_bsbm_7}, if an \texttt{ORDER BY} was present in the SPARQL, it would be represented as $\tau_{column}$ and be the parent of $\pi$. In SPARQL, \texttt{ORDER BY} operates on a variable. Using the same process as for identifying projected columns for the \texttt{SELECT}, we use the subject-triple pattern mapping to identify which Hive column to sort by.

\texttt{FILTER}. We represent the SPARQL regular expression \texttt{FILTER} as a selection $\sigma$ in the AST to conform with standard relational algebra. The regex string is used as the argument for $\sigma$. While in AST form, \texttt{FILTER} by regex is handled the same as a non-regex \texttt{FILTER}. They remain the same until we differentiate the two during generation of HiveQL where the regex filter is emitted as a \texttt{LIKE}.

\texttt{DESCRIBE}. This is equivalent to a \texttt{SELECT ALL} in standard SQL. We represent this in the AST as a projection $\pi$ where we project all columns in our property table. This is an expensive operation since we must access all columns in our Hive table which are scattered throughout HDFS.

\texttt{LIMIT}. This occurs at the end of the SPARQL query. It is not supported by relational algebra and as a consequence, it is not represented in our AST. When generating HiveQL, we must manually check the original SPARQL query for the existence of this operator. This is described in more detail when we generate HiveQL.

\texttt{UNION}. Unions in SPARQL combine two relations or subquery results. Representing this in our AST requires two subqueries, with identically projected columns before the union occurs. To do this, we recurse into each of the SPARQL \texttt{UNION} relations and perform the same process as if we started with a brand new SPARQL query. This will result in two subqueries projecting identical columns. Hive views will be created as necessary by each of the subqueries. The view generation process is identical to generating views in the primary (outer) query.

\texttt{REDUCED}. This operator is used as a hint to the query engine. It indicates that duplicates can be eliminated but it is not required. In our system's query engine, we abide by all \texttt{REDUCED} requests and remove duplicates. It is represented in our AST by a duplicate elimination operator $\delta$. No operands are required.

\texttt{BOUND}. This operator requires a single variable as an operand. As defined in SPARQL, this operator returns true if the variable is bound (i.e. is not null) and false otherwise. The \texttt{BOUND} operator translates to a selection $\sigma$ where the bound variable is not null. To find the specific column to which the variable is referring to, we use the subject-triple pattern mapping shown in Figure \ref{fig:join_keyset}.

The next section outlines how we generate the HiveQL statement using the original SPARQL query and newly generated AST.

\subsection{HiveQL Generation} \label{sec:hql_generation}

The final step is to generate the HiveQL query which will be executed on Hive, HBase, and MapReduce. We have done most of the work and simply need to put it in HiveQL (SQL) form. The remaining steps require specifying which relation the projection attributes will be retrieved from. We will continue with both of our example queries: the simple join query in Codebox \ref{code:sparql_join_query} and the complex Berlin Query 7 in Codebox \ref{code:bsbm_q7_sparql}. We will focus on the join query first.

Using the newly constructed AST in Figure \ref{fig:join_query_ast}, we begin generating HiveQL from the root node. We see the root node is a projection and thus go to the subject-triple pattern mapping in Figure \ref{fig:join_keyset}. From here, we can find the appropriate Hive view where each projection variable is stored. The projection variables for the join query are \texttt{?firstName, ?age} and \texttt{?population}. These refer to columns stored in R1, R1, and R2, respectively. From the subject-triple pattern mapping, we also know the full column name for each of these variables. Converting this into HiveQL, we get the statement: \texttt{SELECT R1.foaf:firstName, R1.foaf:age, R2.foaf:population}. Colons here do not denote the column family but instead refer to the namespaces used in the SPARQL query.

We continue traversing the tree until we encounter a join operator, at which point we explore the left-subtree first. Any selection, $\sigma$, in the left-subtree is included in the join condition of the nearest ancestral join operator. The result of concatenating the selection is shown in line 4 of Codebox \ref{code:join_hiveql}. Had there been additional equality selections in the right-subtree, those selections would be included in the list of join conditions.

\begin{figure}[h]
	\singlespacing
	\begin{mdframed}
		\begin{lstlisting}[label=code:join_hiveql,caption=Berlin SPARQL Benchmark Query 7]
		SELECT
		R1.foaf:firstName, R1.foaf:age, R2.foaf:population
		FROM
		R1 JOIN R2 ON (R1.foaf:livesIn = R2.key AND R1.rdf:type = rdf:Person)
		\end{lstlisting}
	\end{mdframed}
\end{figure}

Berlin Query 7 undergoes the same AST to HiveQL conversion process as the join query example. As a rule of thumb, selections are included as an additional join condition of its nearest ancestral join operator. Selections located at the top of the AST with no ancestral join operator are placed in the \texttt{WHERE} clause of the HiveQL query. All non-equality selections and regular expression filters are also included in the \texttt{WHERE} clause. This is a non-optimal placement since the optimal placement of a selections is before a join operator. However, Hive places this restriction on HiveQL queries.

Regular expression comparisons are denoted in HiveQL by using the \texttt{LIKE} operator. During the AST traversal, any selection containing a reserved regex symbol will be marked as a regex filter. As a result, the HiveQL generator will emit a HiveQL fragment with \texttt{LIKE}. Both non-equality and regex filters are placed in the \texttt{WHERE} clause as an additional condition.

Abstract syntax trees with more than one join are denoted as an n-way join. Joins are appended to the HiveQL query by AST level first, then by left-to-right. The resulting HiveQL for the Berlin Query 7 is shown in Codebox \ref{code:bsbm_q7_hiveql}.

\begin{figure}[h]
	\singlespacing
	\begin{mdframed}
		\begin{lstlisting}[label=code:bsbm_q7_hiveql,caption=Berlin SPARQL Benchmark Query 7 as HiveQL]
		SELECT
		R1.label, R2.key, R2.price, R2.vendor, R3.label, R4.key, R4.title,
		R4.reviewer, R5.name, R4.rating1, R4.rating2
		FROM
		R1 LEFT OUTER JOIN R2 ON (R1.key="%ProdXYZ%" AND R1.key=R2.product)
		JOIN R3 ON (R2.vendor = R3.key)
		LEFT OUTER JOIN R4 ON (R4.key = R1.reviewFor)
		LEFT OUTER JOIN R5 ON (R4.reviewer = R5.key)
		WHERE
		R3.country = "<http://downlode.org/rdf/iso-3166/countries#DE>"
		AND R2.validTo > %currentDate%
		\end{lstlisting}
	\end{mdframed}
\end{figure}

Left outer joins are handled the same way as natural joins. Because of our parsing strategy, full- and right-outer joins do not occur in our AST and will not appear in the HiveQL query.

Unions are placed in the HiveQL \texttt{WHERE} clause with two subqueries. Each subquery is given its own set of Hive views from which the subquery is executed on. The Hive views must remain separate and are assigned different view names in the HiveQL query. The SPARQL \texttt{BOUND} is placed in the HiveQL \texttt{WHERE} clause as an \texttt{IS NOT NULL}. The \texttt{DISTINCT} and \texttt{REDUCED} keywords are placed immediately following \texttt{SELECT} but before the list of projected columns. Any limits or ordering requirements are placed at the end of the HiveQL query.

Once the HiveQL generator has evaluated every node of the AST, the HiveQL query is then constructed as a single string and executed on Hive. Hive generates a physical query plan (see Appendix \ref{sec:appendix_physical_plan1}) and begins execution of the MapReduce jobs. Once all jobs have completed, all materialized views are discarded, the result is returned as XML, and the query is complete.

\section{Experimental Setup} \label{sec:setup}

In our NoSQL system, many techniques were borrowed from relational database management systems. Since the primary aim of this system is to cache the Semantic Web by storing massive amounts of RDF data, we must analyze the performance in a cloud setting. To evaluate this system, we perform several benchmark tests in a distributed environment.
The benchmarks used in the evaluation are described in Section \ref{sec:benchmarks}, the computational environment is outlined in Section \ref{sec:comp_env}, and system settings in Section \ref{sec:system_settings}. Results are presented in Section \ref{sec:results} followed by a discussion in Section \ref{sec:discussion}.

\subsection{Benchmarks} \label{sec:benchmarks}

Two benchmarks were used in the evaluation of the system: the Berlin SPARQL Benchmark (BSBM) and the DBpedia SPARQL Benchmark. Both of these benchmarks are freely available online for download.

\subsubsection{Berlin SPARQL Benchmark}
The Berlin SPARQL Benchmark (BSBM) is built around an e-commerce use case in which a set of products is offered by different vendors while consumers post reviews about the products \cite{bizer2009berlin}. The benchmark query mix emulates the search and decision making patterns of a consumer exploring products to purchase. The BSBM dataset is synthetic and was generated using three scaling factors:

\begin{itemize}
	\singlespacing
	\item Scale Factor: 28,850 resulting in 10,225,034 triples (10 million)
	\item Scale Factor: 284,826 resulting in 100,000,748 triples (100 million)
	\item Scale Factor: 2,878,260 resulting in 1,008,396,956 triples (1 billion)
\end{itemize}

The file sizes for the BSBM 10 million, 100 million, and 1 billion triples datasets are roughly 2.5 GB, 25 GB, and 250 GB, respectively.

\subsubsection{DBPedia SPARQL Benchmark}

The DBpedia SPARQL Benchmark is based on real-world queries executed by humans on DBpedia\footnote{\url{http://www.dbpedia.org}} \cite{morsey2011dbpedia}. We use the dataset generated from the original DBpedia 3.5.1 with a scale factor of 100\% consisting of 153,737,783 triples. This dataset is 25 GB in size.

\subsection{Computational Environment} \label{sec:comp_env}

All experiments were performed on the Amazon Web Services EC2 Elastic Compute Cloud\footnote{\url{http://aws.amazon.com}} and Elastic MapReduce service. All cluster nodes were m1.large instances with the following specifications:

\begin{itemize}
	\singlespacing
	\item 8 GB (7.5 GiB) main memory
	\item 840 GB (2x420 GB) of local disk storage
	\item Intel Core 2 Duo T6400 @ 2.00 GHz
	\item 4 Elastic Compute Units (ECU) comprised of 2 virtual CPUs
	\item 64 Bit Architecture Intel Processors
\end{itemize}

A virtual CPU is equivalent to a physical core. Elastic Compute Units (ECU) measure the speed of each machine. One ECU is the equivalent CPU capacity of a 1.0-1.2 GHz 2007 Opteron or 2007 Xeon processor or an early-2006 1.7 GHz Xeon processor. Amazon Elastic MapReduce was used to setup and configure the Hadoop/HBase cluster. One master node is used in all configurations. The number of slave nodes range from 1 to 16 in powers of two. Specific instructions on how to reproduce the evaluation environment are listed in Appendix \ref{appendix_reproduce_emr}.

To aid in reproducibility and comparability, we ran Hadoop's TeraSort \cite{o2008terabyte} on a cluster consisting of 16 m1.large EC2 nodes (17 including the master). Using TeraGen, 1 TB of data was generated in 3,933 seconds (1.09 hours). The data consisted of 10 billion, 100 byte records. The TeraSort benchmark completed in 11,234 seconds (3.12 hours) \cite{haque2013nosqlrdf}. Detailed configuration settings for TeraSort can be found in Appendix \ref{appendix_load_terasort}.

We evaluated our system on 1, 2, 4, 8, and 16 node clusters. An additional master node was used for various housekeeping tasks such as Apache Zookeeper\footnote{\url{http://zookeeper.apache.org}}. Any dataset which took longer than 24 hours to load was not evaluated. For each benchmark run, we considered two metrics: the arithmetic mean and the geometric mean. The geometric mean dampens the effect of outliers -- which we use for all plots and discussions.

\subsection{System Settings} \label{sec:system_settings}

Appendix \ref{appendix_hbase_paramaters} lists HBase parameters that were changed from their default value. A single column family was used to store all predicate columns.

The specific software packages used were: Hadoop 1.0.3, HBase 0.92, and Hive 0.8.1. One zookeeper was running on the master for all cluster configurations. Each HBase RegionServer was given 5 GB of memory while the rest was allocated to Hadoop. All nodes were located in the North Virginia region. Apache Jena 2.7.4 was used for the query layer.

\section{Results} \label{sec:results}

\begin{figure}[t]
	\centering
	\caption{Database Loading Times}
	\includegraphics[width=.8\textwidth]{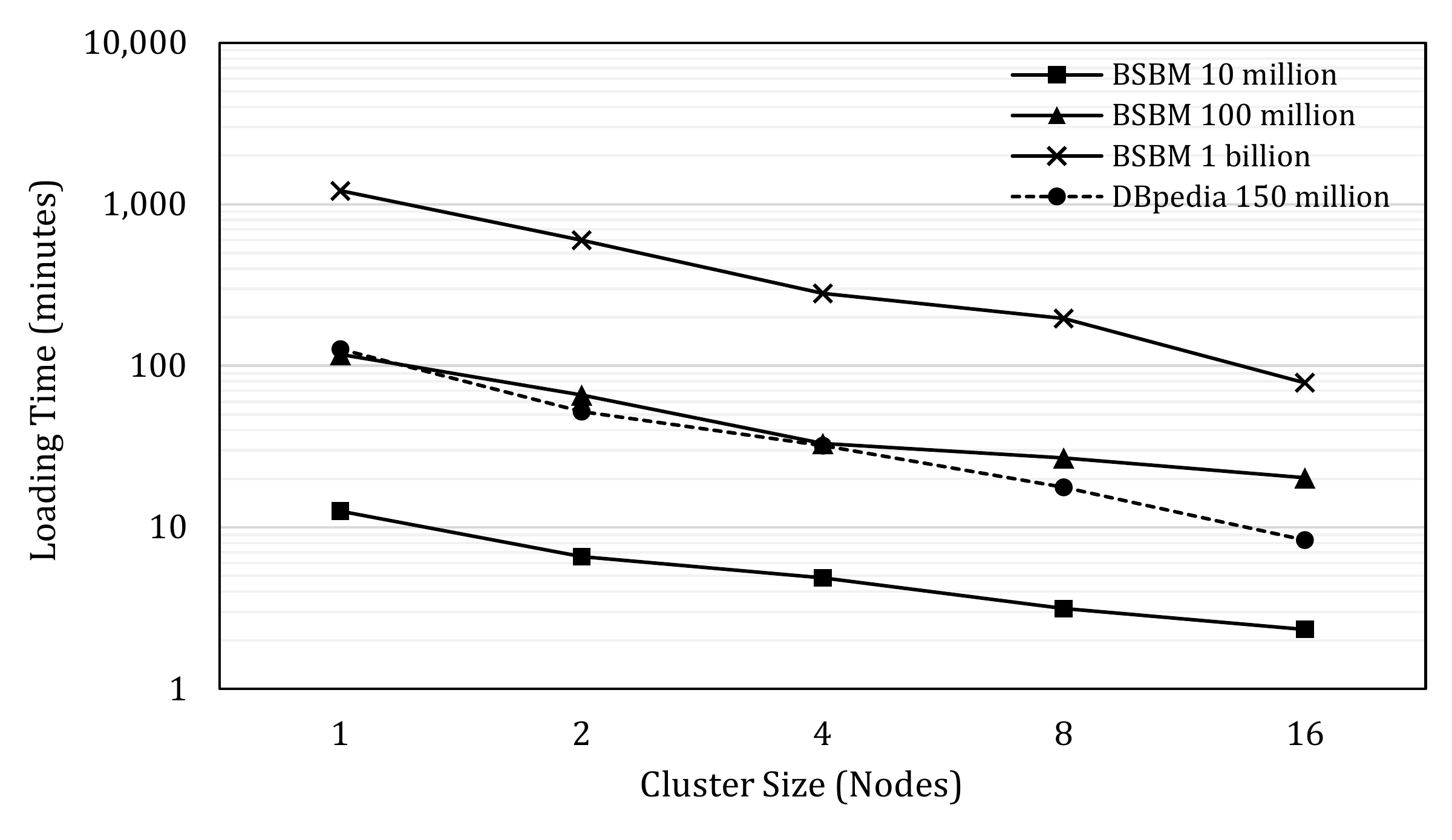}
	\label{fig:loadtimes}
\end{figure}

We now describe the results of the evaluation. Section \ref{sec:load_times} describes the loading times for each cluster and benchmark. Section \ref{sec:qet} provides query execution times for the BSBM and DBpedia datasets.

\subsection{Loading Times} \label{sec:load_times}

Loading time excludes the time spent moving the file from the local machine into HDFS. We define loading time as the time spent performing all necessary transformations on the dataset plus time spent inserting into the database (converting into HBase's internal format). The loading times for both BSBM and DBpedia are shown in Figure \ref{fig:loadtimes}. We implement a single MapReduce program to transform and populate the database. As a result, as the number of nodes increases, the load time decreases at a linear rate.

\subsection{Query Execution Times}\label{sec:qet}

\begin{figure}[!h]
	\begin{minipage}{\textwidth}
		\centering
		\caption{Query Execution Time: BSBM 10 Million Triples}
		\includegraphics[width=\textwidth]{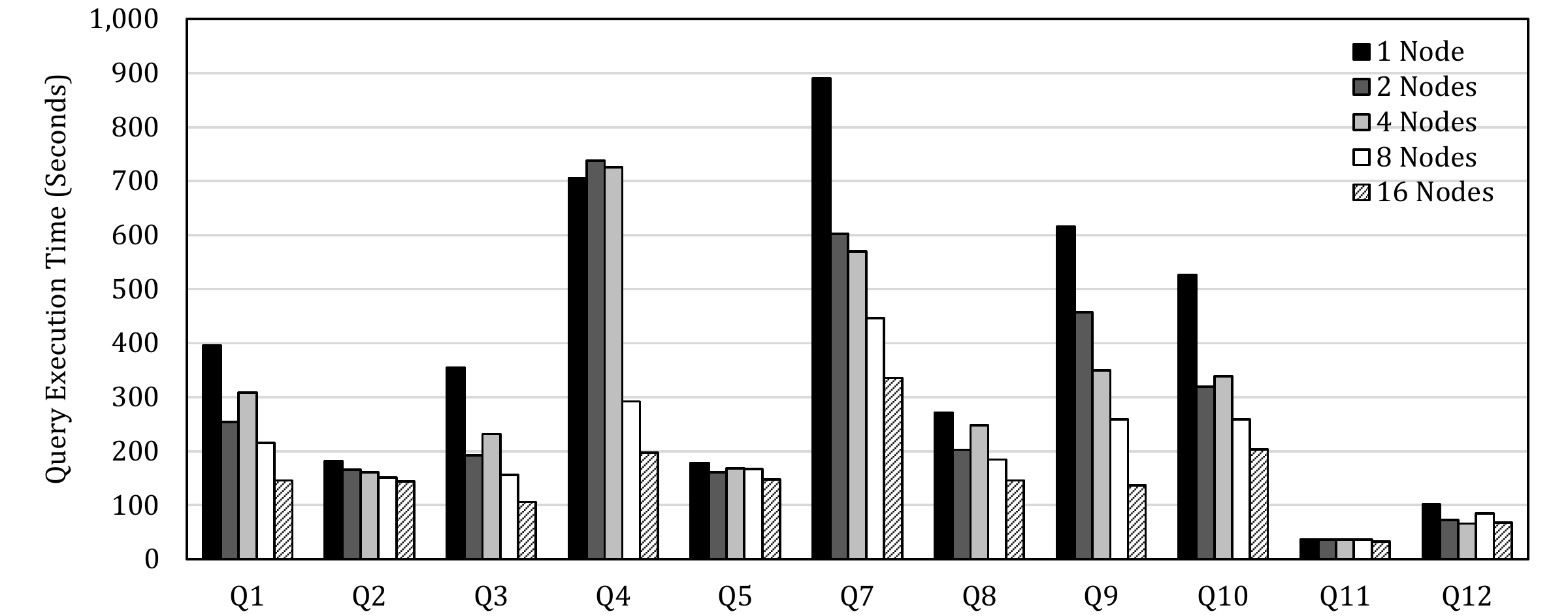}
		\label{fig:qet_bsbm10m}
		\caption{Query Execution Time: BSBM 100 Million Triples}
		\includegraphics[width=\textwidth]{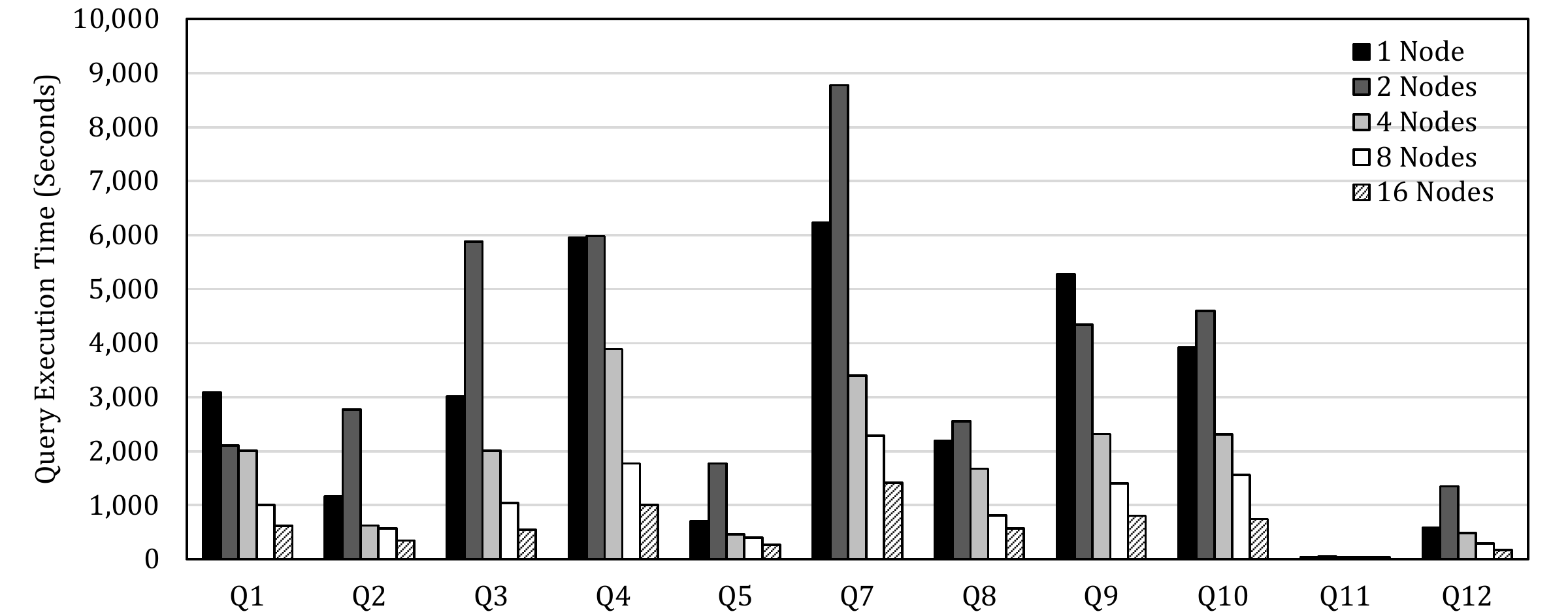}
		\label{fig:qet_bsbm100m}
	\end{minipage}
\end{figure}

\begin{figure}[!h]
	\begin{minipage}{\textwidth}
		\centering
		\caption{Query Execution Time: BSBM 1 Billion Triples}
		\includegraphics[width=\textwidth]{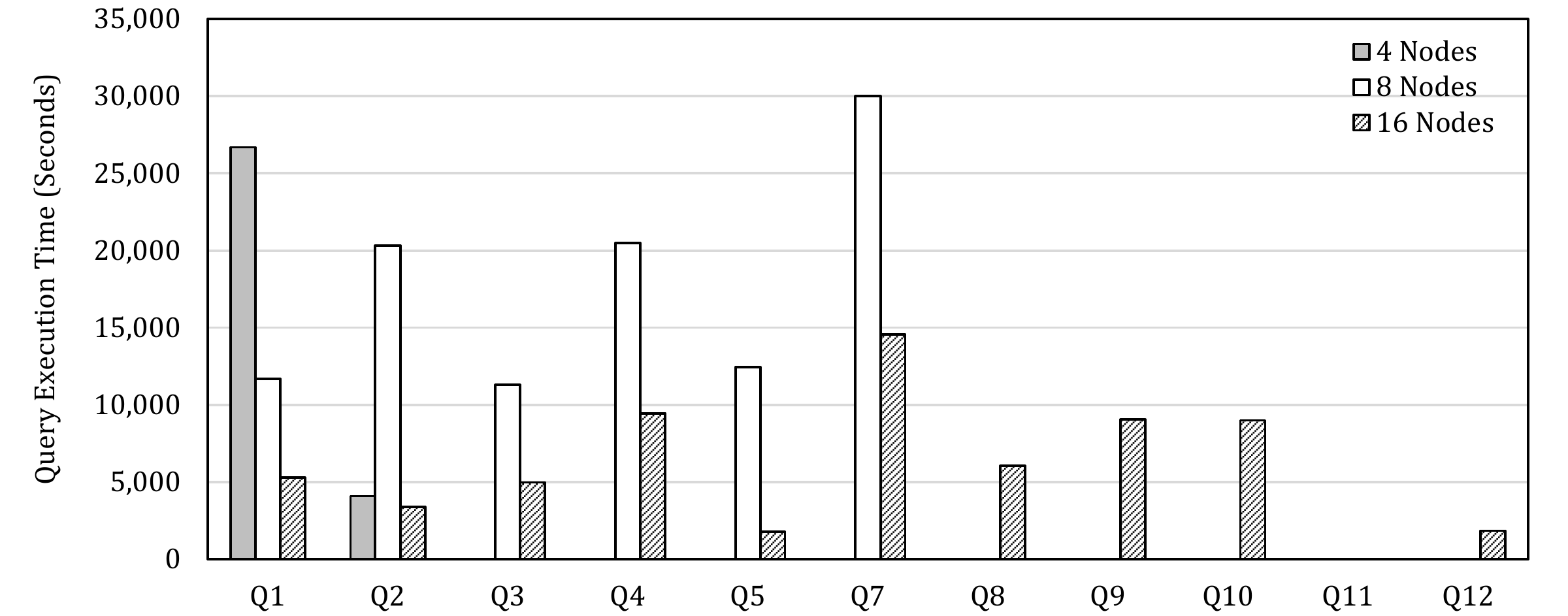}
		\label{fig:qet_bsbm1000m}
		\centering
		\caption{Query Execution Time: DBpedia 150 Million Triples}
		\includegraphics[width=\textwidth]{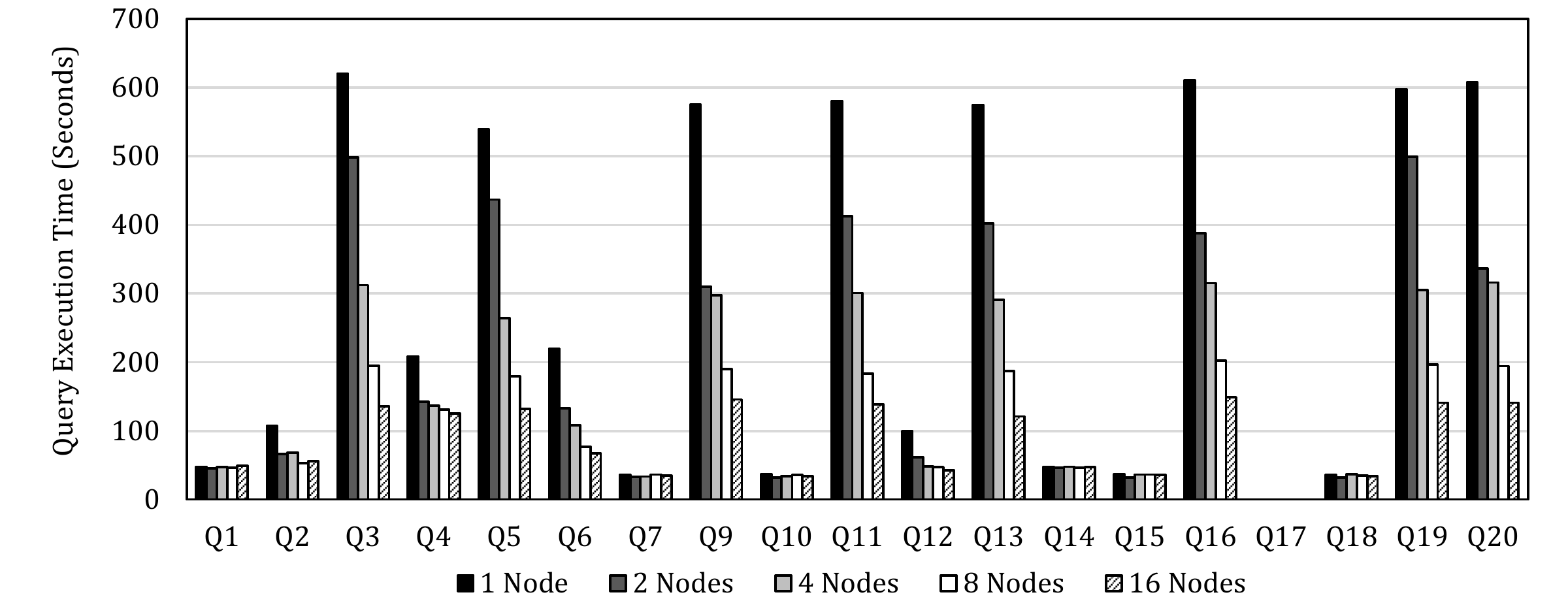}
		\label{fig:qet_dbp150M}
	\end{minipage}
\end{figure}

We show the plots for the four different datasets and sizes. The Berlin Benchmark excludes Query 6 from the official test driver so it is omitted from our plots. Figures \ref{fig:qet_bsbm10m}, \ref{fig:qet_bsbm100m}, and \ref{fig:qet_bsbm1000m} show the results for the Berlin SPARQL Benchmark. Exact evaluation numbers can be found in Appendix \ref{appendix_qet}.

Note that in Figure \ref{fig:qet_bsbm100m}, Query 7 requires the most time to complete. This can be attributed to its 5-way join. When the dataset is increased from 100 million to 1 billion triples, some queries begin to take prohibitively long. As shown in Figure \ref{fig:qet_bsbm1000m}, the Berlin Benchmark consisting of 1 billion triples caused queries to run for several hours. Figure \ref{fig:qet_dbp150M} shows the results for the DBpedia SPARQL Benchmark.

\section{Discussion} \label{sec:discussion}
\subsection{Dataset Characteristics}
\begin{figure}[!h]
	\centering
	\caption{Dataset Characteristics: Node Degree Frequency}
	\begin{subfigure}[h]{\textwidth}
		\caption{BSBM (100 Million triples)}
		\includegraphics[width=\textwidth]{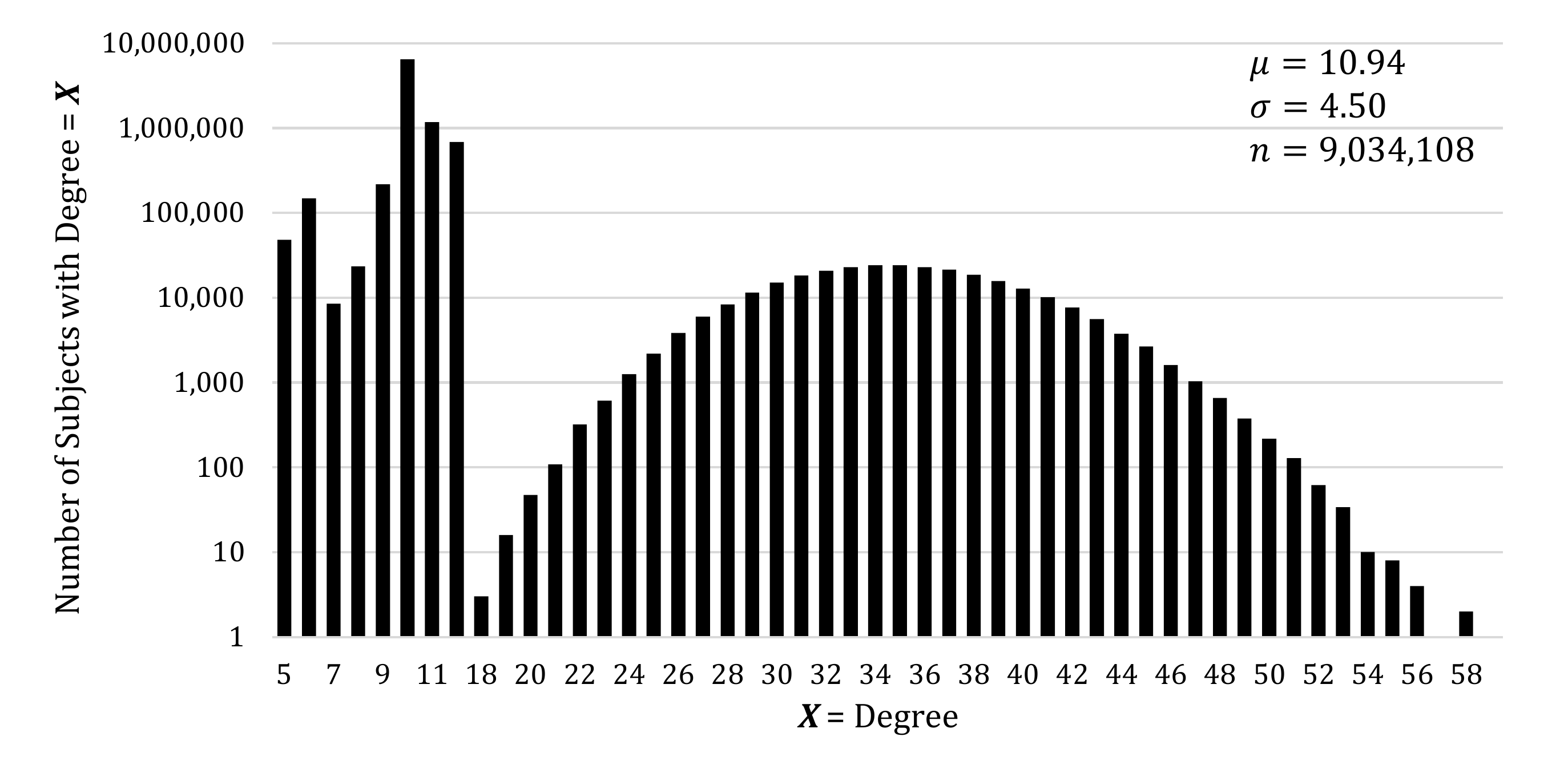}
		\label{fig:bsbm100m_degree}
	\end{subfigure}
	\begin{subfigure}[h]{\textwidth}
		\caption{DBpedia (150 Million triples)}
		\includegraphics[width=\textwidth]{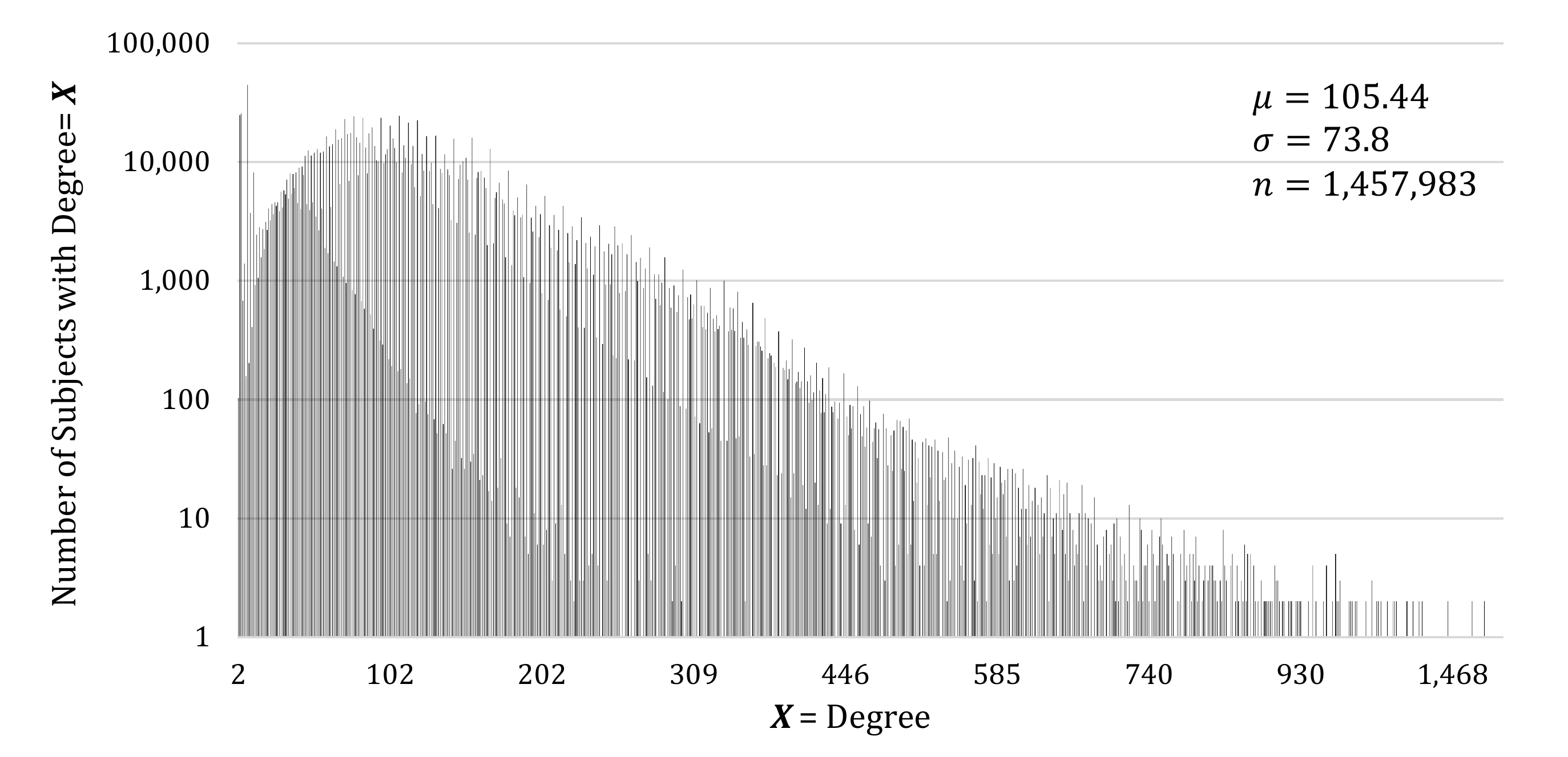}
		\label{fig:dbpedia_degree}
	\end{subfigure}
\end{figure}

We performed an analysis of the datasets to better understand our systems strengths and weaknesses as well as possible causes. The first metric we calculate is the average node degree, that is the number of triples associated with a specific subject. The analysis was performed using two MapReduce jobs. The first job identified unique subjects and the number of objects adjacent with each subject. The second MapReduce job consolidated the node degree counts into a set of frequencies. The results are shown in Figures \ref{fig:bsbm100m_degree} and \ref{fig:dbpedia_degree}.

The analysis of the BSBM dataset was done on the 100 million triples dataset and the DBpedia dataset contained 150 million triples. Since the BSBM dataset is synthetic, both the 10 million and 1 billion triples dataset follow the same frequency distribution. The number of unique predicates, represented as columns in our table, is 40 for BSBM and 23,343 for DBpedia. Given these values, we can conclude that DBpedia contains very sparse data while the BSBM dataset is more densely populated. Because of this, the DBpedia dataset makes better use of HBase's null compression than BSBM.

Given any random subject $S^*$ from the BSBM dataset, on average, $S^*$ will contain object values for 28\% of the all columns. This is excludes multi-valued attributes and hence is an upper bound on the expectation. Contrast this with DBpedia, where each subject, on average, contains values for 0.4\% of all columns. We will now examine the query results for each of the datasets.

\subsection{Berlin SPARQL Benchmark}

As shown in Table \ref{bsbm_query_characteristics}, the BSBM queries use a diverse set of SPARQL query operators. These queries generally touch a lot of data with variable subjects. The only SPARQL and RDF features excluded by this benchmark are named graphs, blank nodes, collections, \texttt{REDUCED} operator, property hierarchies, and queries in the \texttt{ASK} form \cite{bizer2009berlin}.

\begin{table}[t]
	\caption[BSBM Query Characteristics]{BSBM Query Characteristics \cite{bizer2009berlin}}
	\includegraphics[width=\textwidth]{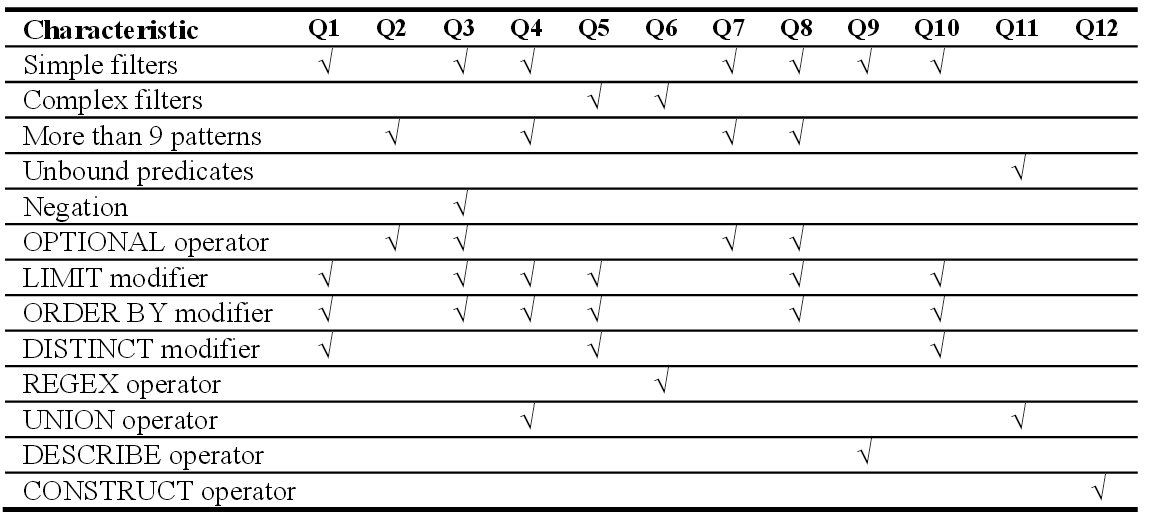}
	\label{bsbm_query_characteristics}
\end{table}

\begin{table}
	\centering
	\caption{Number of Joins Required by BSBM Queries}
	\begin{tabular}{|l|c|c|c|c|c|c|c|c|c|c|c|}
		\hline
		\textbf{BSBM Query}	 	 & 1 & 2 & 3 & 4 & 5 & 7 & 8 & 9 & 10 & 11 & 12 \\ \hline
		\textbf{Number of Joins} & 0 & 2 & 0 & 0 & 1 & 4 & 1 & 1 & 1  & 0  & 0 \\ \hline
	\end{tabular}
\end{table}

The slowest BSBM queries are queries 4, 7, and 9. Query 4 uses the \texttt{UNION} operator. Queries 7 and 9 require 4 and 1 join, respectively. Note that the number of relations being joined is the number of joins plus one. For the 1 billion triples dataset, the number of rows with a join key satisfying the join conditions becomes so large that the queries begin to take greater than one hour. The only queries that are able to complete in less than one hour are queries 2, 5, 11, and 12. If we take a closer look at these shorter queries, query 2 immediately restricts the rows being joined by specifying the key as a constant. As a result, fewer rows are being sent across the network. Query 5 requires one join but also places a restriction on the join key, both on Relation 1 (R1) and Relation 2 (R2), where $R1 \bowtie R2$. Both queries 11 and 12 also place restrictions on Relation 1 before performing any joins. Variable predicates are currently not supported by the AST builder and HiveQL generator. As a result, BSBM Query 11 requires a manual translation into Hive code. Using the manually translated HiveQL, the query is a projection of several variables with a single equality  condition in the \texttt{WHERE} clause and requires no joins. As a result, Query 11 experiences fast execution times on all dataset sizes.

\begin{table}
	\centering
	\caption[BSBM Query Selectivity]{BSBM Query Selectivity \cite{sequeda2012ultrawrap}}
	\begin{tabular}{|l|c|c|c|}
		\hline
		& Inner Join & Left-Outer Join & Variable Predicate \\ \hline
		\textbf{Low Selectivity} & 1, 3, 10 & & \multirow{2}{*}{9, 11} \\
		\cline{1-3} \textbf{High Selectivity} &4, 5, 12 & 2, 7, 8 &  \\ \hline
	\end{tabular}
\end{table}

The effect of bloom filters will be dictated by the query selectivity. Queries with high selectivity such as 5, 8 and 12, touch fewer data and generally have faster runtimes than low selectivity queries. When combined with joins, high selectivity queries that require specific rows will take full advantage of bloom filters and reduce the number of blocks read. This does not apply to query 7 because the overhead due to the 5-way join makes any gains due to bloom filters insignificant. Additionally, queries with variable predicates are not fully supported by the system. As a result, these queries are unable to execute for larger datasets.

\subsection{DBpedia SPARQL Benchmark}

The DBpedia SPARQL Benchmark introduces simpler queries on a more sparse dataset. The results presented in Figure \ref{fig:qet_dbp150M} indicate faster runtimes for the DBpedia dataset than the BSBM dataset of similar size (BSBM 100 million in Figure \ref{fig:qet_bsbm100m}). Looking at Figure \ref{fig:qet_dbp150M}, we can group queries into two categories: (i) queries that initially require long runtimes and decrease as the cluster size increases and (ii) queries that require relatively (fast) constant time for all cluster sizes. The first group of queries that require long runtimes initially are: 3, 5, 9, 11, 13, 16, 19, and 20. The characteristics of these queries are as follows:
\begin{itemize}
	\item Queries with string operations on a variable subject or object will require longer time. This includes language filtering and substring searching.
	\item Queries with one or more joins require multiple MapReduce jobs and will require additional time. This includes \texttt{OPTIONAL} clauses which are translated into left outer joins.
	\item Queries with variable subjects in the \texttt{WHERE} clause tend to perform more slowly.
\end{itemize}

\noindent
Queries that execute in near-constant time for all cluster sizes are reaching a minimum query execution time. This can be attributed to the overhead associated with generating the MapReduce jobs.
The variance between runtimes for constant-time queries can be attributed to various factors including Amazon network usage spikes or delays due to abnormal behavior of mapper or reducer nodes. Below, we list characteristics shared by the DBpedia queries that execute between 30 and 60 seconds:

\begin{itemize}
	\item Subjects are specified and thus we can take full advantage of bloom filters for fast row access.
	\item Queries generally do not involve variable subjects or objects.
	\item Unions are not complex (e.g. no joins).
	\item Queries do not perform string or filter operations. Conditionals such as \texttt{NOT NULL} execute faster than conditionals involving equalities or inequalities.
\end{itemize}

\noindent
Query 17 reports no results since the SPARQL query requires logical operators which is currently unsupported by our parser.

\subsection{Joins}

Excluding the size of the dataset, the most important factor in dictating query execution times is the number of joins. Let $|P|$ denote the number of columns in the property table, let $|S|$ denote the number of unique subjects and let $n$ denote the number of joins. In the worst case, an n-way join will require reading $(|S||P|*n)$ triples from HDFS, sorting these values, and transmitting them across the cluster. This estimated excludes multivalued attributes. As we can see here, great care must be taken to eliminate rows that are unlikely to be joined.

\subsection{Node Activity}

\begin{figure}[t]
	\caption{Cluster Activity -- BSBM 10 Million Triples on 16 Nodes}
	\makebox[\linewidth][c]{
		\hspace{2 mm}
		\begin{subfigure}[b]{.6\textwidth}
			\centering
			\caption{CPU Utilization}
			\includegraphics[width=.95\textwidth]{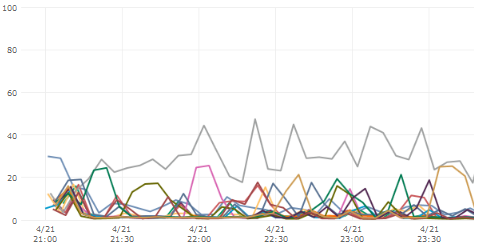}
			\label{fig:16n_10m_cpu}
		\end{subfigure}%
		\begin{subfigure}[b]{.6\textwidth}
			\centering
			\caption{Network Out}
			\includegraphics[width=.95\textwidth]{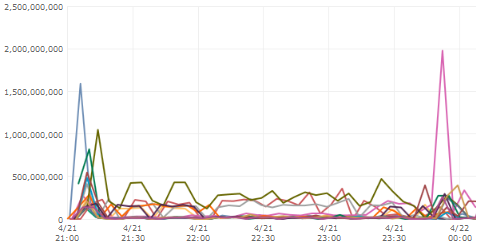}
			\label{fig:16n_10m_net_out}
		\end{subfigure}%
	}\\
	\label{fig:cluster_activity_10m_16n}
\end{figure}

Join selectivity and the number of joins are responsible for the majority of network traffic in the cluster. In Figure \ref{fig:cluster_activity_10m_16n}, we show data collected during the Berlin SPARQL Benchmark runs for the master and several slave nodes. All graphs depict metrics for 1 minute intervals. CPU utilization is averaged over 1 minute and network or disk is the total bytes moved per minute.

From Figure \ref{fig:16n_10m_cpu}, we know that all cluster nodes stay beneath 50\% CPU utilization. Almost all CPU activity will be for equality comparisons (of joins) or the MapReduce sort phase. Since the dataset is small (2.5 GB), the CPUs do not have large workloads for sorting and/or perfoming comparisons. No machine fully utilizes any single CPU core. Figure \ref{fig:16n_10m_net_out} shows the outgoing network traffic for each of the cluster nodes. Since each machine is equipped with 8 GB of memory, they can fit the entire dataset in memory -- although in most cases it is not necessary to scan the entire dataset. As a result, nodes can cache HDFS blocks in memory and do not have to request blocks from other nodes. However, this does not eliminate all network traffic as blocks must be occasionally sent to other mappers/reducers.

\begin{figure}[t]
	\caption{Cluster Activity -- BSBM 1 Billion Triples on 16 Nodes}
	\vspace{3 mm}
	\makebox[\linewidth][c]{%
		\begin{subfigure}[b]{.6\textwidth}
			\centering
			\caption{CPU Utilization}
			\includegraphics[width=.95\textwidth]{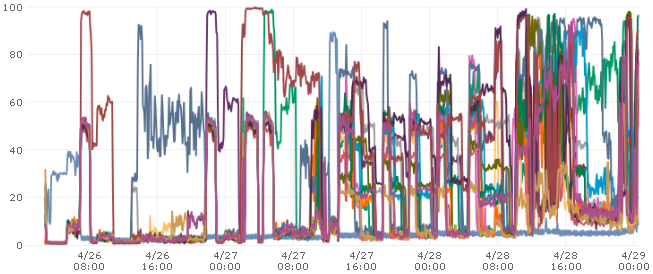}
			\label{fig:16n_1b_cpu}
		\end{subfigure}%
		\begin{subfigure}[b]{.6\textwidth}
			\centering
			\caption{Network Out}
			\includegraphics[width=.95\textwidth]{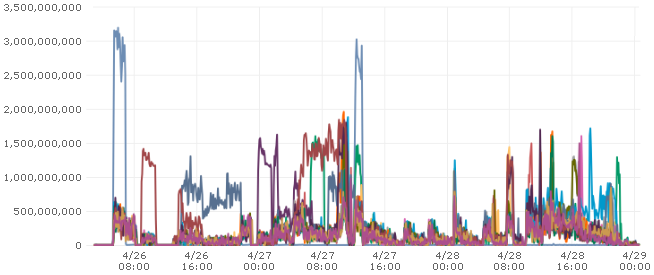}
			\label{fig:16n_1b_net_out}
		\end{subfigure}%
	}\\
	\makebox[\linewidth][c]{%
		\begin{subfigure}[b]{.6\textwidth}
			\vspace{5 mm}
			\centering
			\caption{Disk Reads}
			\includegraphics[width=.95\textwidth]{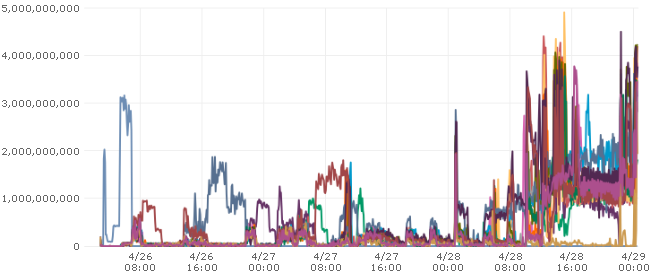}
			\label{fig:16n_1b_disk_reads}
		\end{subfigure}%
		\begin{subfigure}[b]{.6\textwidth}
			\centering
			\caption{Live Map Tasks}
			\includegraphics[width=.95\textwidth]{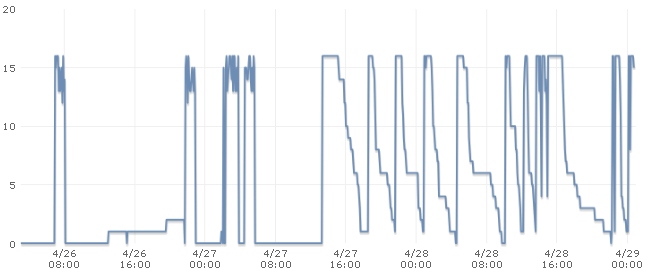}
			\label{fig:16n_1b_map_tasks}
		\end{subfigure}%
	}
	\label{fig:16n_1b_bsbm}
\end{figure}

Contrast the cluster activity of BSBM 10 million triples with the cluster activity of BSBM 1 billion triples shown in Figure \ref{fig:16n_1b_bsbm}. We will look at four metrics: CPU utilization, network traffic out of each node, disk reads, and the number of map tasks.

Because each node now has larger map and reduce assignments, the CPU must be constantly processing the input. The 10 million BSBM dataset is 2.5 GB in size. The 1 billion triples dataset is 250 GB. The effect of this is apparent in Figure \ref{fig:16n_1b_cpu}. Nearly every node performs at 50\% CPU utilization or higher. There is less idling since each worker has a longer task queue. These tasks are typically equality comparisons for the join condition of \texttt{WHERE} clauses or the tasks are sorting the mapper output.

Moving to Figure \ref{fig:16n_1b_net_out}, we can see that all cluster nodes move information from their local disks to other nodes. This is expected due to the nature of HDFS's sharding and replication. However, some nodes exhibit network-out rates of 1 GB or higher for several minutes -- sometimes sending data out for over an hour. All data associated with an RDF subject is typically stored on the same HBase RegionServer. Depending on the characteristics of the query, if a query references many columns in a Hive table, the specific RegionServer/slave housing the RDF subject will be required to send all its data for that subject across network. This results in one node transmitting more data than other nodes and thus explains the cases of large network out.

The disk reads in Figure \ref{fig:16n_1b_disk_reads} are correlated with the network I/O. However, additional disk reads may be necessary when the relation cannot fit in memory. On the far right of Figure \ref{fig:16n_1b_disk_reads}, most cluster nodes experience 1.5 GB of bytes reads from disk per minute (about 25 MB/sec). Figure \ref{fig:16n_1b_map_tasks} shows us the number of currently running map tasks. In the left half of the graph, all mappers finished their tasks in roughly the same time. However, on the right side, most likely attributed to data skew, some mappers lag behind others.

Summarizing the experiment and analysis -- queries with unfiltered joins require larger amounts of data to be sent across the network resulting in slow query runtimes. Reducer nodes should be given larger amounts of memory compared to their mapper counterparts. Mappers must read from disk regardless of the MapReduce step. However, reducer nodes can place incoming data from the network directly into memory. Additional memory would be advantageous for reducer nodes. Bloom filters reduce the number of HDFS blocks read during table scans. As a result, sparse datasets are able to benefit from this feature and thus enjoy shorter query runtimes. However, DBpedia also contains simpler benchmark queries.

\section{Related Work} \label{sec:related}

Currently a highly debated topic, both classical relational databases and MapReduce databases both have their advocates and critics. Most enterprises and startups today are flocking towards distributed database systems due to the era of ``big data". MapReduce has been proven to scale efficiently, reduce workload execution times, and enable more powerful analyses.

However, MapReduce is not suitable for all workloads. Several studies have concluded that for datasets less than 1 terabyte (TB), parallel database systems either perform on par or outperform MapReduce on a cluster of 100 nodes \cite{Stonebraker:2010:MPD:1629175.1629197, Pavlo:2009:CAL:1559845.1559865}. Several factors affect the performance of MapReduce that do not apply to parallel databases \cite{Jiang:2010:PMI:1920841.1920903}. Despite these results, several techniques and systems have been created to solve large-scale data warehousing problems.

Several cloud triple stores exist today and have vast differences among one another. Among these systems are 4store \cite{harris20094store}, CumulusRDF (Cassandra) \cite{ladwig2011cumulusrdf}, HBase \cite{sever_thesis}, and Couchbase \footnote{\url{http://www.couchbase.com}}. In a comparative study of all of these triple-stores (including our HBase-Hive implementation), systems using MapReduce introduced significant query overhead while strictly in-memory stores were unable to accommodate large datasets \cite{haque2013nosqlrdf}. It was discovered that queries involving complex filters generally performed poorly on NoSQL systems. However, traditional relational database query optimization techniques work well in NoSQL environments.

It is clear that the increasing amount of data, both structured and unstructured, will require storage in a distributed manner. Many systems using Google BigTable variants such as Rya \cite{punnoose2012rya}, Jena HBase \cite{khadilkar2012jena}, and others \cite{sever_thesis, husain2009storage} are able to efficiently scale horizontally but often execute queries ranging from seconds to hours. If we are to migrate future data systems to a distributed environment, several improvements at the data and query layer are needed.

\section{Conclusion} \label{sec:conclusion}

In this thesis, we explored the fundamentals our system's underlying components and their intended purpose in our overall database system. We then proceeded to describe our property table data model and advantages it offers us. The query layer consists of identifying joins given a SPARQL query and generating temporary views to in an attempt to minimize data transfer during MapReduce.

The results show that sparse datasets make use of many of our property table design features such as bloom filters and self-join eliminations to improve query performance. For larger workloads and denser graph datasets, performance becomes dependent on the join algorithm implementation. SPARQL queries with n-way joins can take hours to complete.

In the future, we plan to improve upon the naive join algorithms implemented by our system by investigating alternative parallel join algorithms. We would also like to provide support for additional SPARQL operators and improve both the AST and HiveQL optimization. Finally, we would like to explore alternatives to Hadoop such as distributed shared-memory systems \cite{engle2012shark}.

\newpage
\bibliographystyle{abbrv}
\bibliography{Thesis}

\newpage
\appendix

\section{Reproducing the Evaluation Environment}
\subsection{Amazon Elastic MapReduce}\label{appendix_reproduce_emr}
To set up the EC2 cluster with Hadoop, HBase, and Hive, use the instructions below and on the project's website\footnote{\url{https://github.com/ahaque/hive-hbase-rdf/}}. These instructions were made using the EC2 portal as of December 2013.

\begin{enumerate}
	\singlespacing
	\item Navigate to the AWS Management Console and go to the \textit{Elastic MapReduce} service.
	\item Click \textit{Create Cluster}.
	\item Under \textit{Software Configuration}, select the appropriate version of Hadoop.
	\item If HBase is not listed under \textit{Applications to be installed}, add it and ignore the backup steps.
	\item If Hive is not listed under \textit{Applications to be installed}, add it and select the appropriate Hive version.
	\item All instance types should be m1.large by default.
	\item Under \textit{Hardware Configuration}, in the \textit{Core} instance type, enter the number of slave nodes for the cluster (1, 2, 4, 8, or 16 used in this experiment).
	\item \textit{Task Instance Group} should be zero.
	\item Select your EC2 key pair and leave all other options to default.
	\item Under \textit{Bootstrap Actions}, add a new, \textit{Custom action}.
	\item For the \textit{S3 Location} enter: \\ \texttt{s3://us-east-1.elasticmapreduce/bootstrap-actions/configure-hbase}.
	\item For ‘Optional Arguments’ enter: \\ \texttt{-s hbase.hregion.max.filesize=10737418240}.
	\item Add the bootstrap action.
	\item Review your settings and click \textit{Create Cluster}.
	\item The cluster will take 3-5 minutes to fully initialize.
\end{enumerate}

\subsection{HBase Parameter Settings} \label{appendix_hbase_paramaters}

Table \ref{hbase_param_table} lists specific parameters and values that were changed in HBase. Many of these settings can be configured in HBase's configuration files or by using the Java API when creating the table.

\begin{table}[H]
	\singlespacing
	\centering
	\caption{Custom HBase Parameters}
	\begin{tabular}{|l|l|}
		\hline
		\textbf{Parameter Name} & \textbf{Value} \\ \hline
		\multicolumn{2}{|l|}{HBase Table} \\ \hline
		hbase.hregion.max.filesize & 10737418240 \\
		hbase.hstore.blockingStoreFiles & 25 \\
		hbase.hregion.memstore.block.multiplier & 8 \\
		hbase.regionserver.handler.count & 30 \\
		hbase.regions.percheckin & 30 \\
		hbase.regionserver.global.memstore.upperLimit & 0.3 \\
		hbase.regionserver.global.memstore.lowerLimit & 0.15 \\ \hline
		\multicolumn{2}{|l|}{Column Family} \\ \hline
		HColumnDescriptor.BloomFilterType & rowcol \\
		HColumnDescriptor.Cache\_Blooms\_On\_Write & true \\
		HColumnDescriptor.MaxVersions & 100 \\ \hline
	\end{tabular}
	\label{hbase_param_table}
\end{table}

\newpage
\section{Detailed Evaluation Tables}
The value NaN denotes the query encountered an error and was aborted or was interrupted due to long runtime.

\subsection{Query Execution Time} \label{appendix_qet}
\singlespacing
\begin{table}[H]
	\caption[Query Execution Time: BSBM 10 Million Triples]{BSBM 10 Million Triples (Geometric Average in Seconds)}
	\centering
	\begin{tabular}{|l|r|r|r|r|r|}
		\hline
		&1 Node&2 Nodes&4 Nodes&8 Nodes&16 Nodes\\ \hline
		Q1&395.82&254.49&307.9&215.35&146.35\\ \hline
		Q2&181.91&165.32&160.97&150.87&144.49\\ \hline
		Q3&354.55&192.83&231.96&156.26&106.05\\ \hline
		Q4&705.96&737.82&725.21&292.48&196.56\\ \hline
		Q5&178.55&161.36&168.21&167.29&147.78\\ \hline
		Q7&889.83&602.84&569.87&446.54&335.63\\ \hline
		Q8&271.94&202.84&247.64&184.74&146.29\\ \hline
		Q9&615.42&457.34&349.46&259.7&137.27\\ \hline
		Q10&526.03&319.47&339.01&258.55&203.17\\ \hline
		Q11&36.64&36.3&36.24&36.52&33.58\\ \hline
		Q12&102.55&72.25&66.01&85.13&68.13\\ \hline
	\end{tabular}
\end{table}

\begin{table}[H]
	\caption[Query Execution Time: BSBM 100 Million Triples]{BSBM 100 Million Triples (Geometric Average in Seconds)}
	\centering
	\begin{tabular}{|l|r|r|r|r|r|}
		\hline
		&1 Node&2 Nodes&4 Nodes&8 Nodes&16 Nodes\\ \hline
		Q1&3085.51&2108.00&2015.53&1003.02&615.66\\ \hline
		Q2&1161.68&2772.77&626.99&569.72&348.65\\ \hline
		Q3&3007.65&5883.78&2014.42&1039.08&544.22\\ \hline
		Q4&5954&5985.29&3888.06&1775.04&1005.08\\ \hline
		Q5&703.21&1771.99&466.63&402.10&270.69\\ \hline
		Q7&6233.79&8774.58&3400.93&2286.69&1417.31\\ \hline
		Q8&2195.31&2559.48&1679.84&813.22&569.76\\ \hline
		Q9&5278.52&4349.19&2321.37&1403.70&805.56\\ \hline
		Q10&3914.52&4603.91&2311.9&1564.91&749.63\\ \hline
		Q11&36.30&50.60&35.94&35.11&35.45\\ \hline
		Q12&582.66&1352.29&484.57&300.05&174.24\\ \hline
	\end{tabular}
\end{table}

\begin{table}[H]
	\caption[Query Execution Time: BSBM 1 Billion Triples]{BSBM 1 Billion Triples (Geometric Average in Seconds)}
	\centering
	\begin{tabular}{|l|r|r|r|r|r|}
		\hline
		&1 Node&2 Nodes&4 Nodes&8 Nodes&16 Nodes\\ \hline
		Q1&NaN&NaN&26696.25&11690.29&5295.24\\ \hline
		Q2&NaN&NaN&40100.8&20337.06&3409.26\\ \hline
		Q3&NaN&NaN&NaN&11289.52&4986.74\\ \hline
		Q4&NaN&NaN&NaN&20473.61&9460.20\\ \hline
		Q5&NaN&NaN&NaN&12458.17&1767.69\\ \hline
		Q7&NaN&NaN&NaN&29989.26&14556.21\\ \hline
		Q8&NaN&NaN&NaN&NaN&6058.42\\ \hline
		Q9&NaN&NaN&NaN&NaN&9060.93\\ \hline
		Q10&NaN&NaN&NaN&NaN&9007.57\\ \hline
		Q11&NaN&NaN&NaN&NaN&36.86\\ \hline
		Q12&NaN&NaN&NaN&NaN&1851.97\\ \hline
	\end{tabular}
\end{table}

\begin{table}[H]
	\caption[Query Execution Time: DBpedia 150 Million Triples]{DBpedia 150 Million Triples (Geometric Average in Seconds)}
	\centering
	\begin{tabular}{|l|r|r|r|r|r|}
		\hline
		&1 Node&2 Nodes&4 Nodes&8 Nodes&16 Nodes\\ \hline
		Q1&47.71&45.32&47.41&46.83&49.1\\ \hline
		Q2&107.05&65.95&68.29&53.16&55.93\\ \hline
		Q3&620.57&498.45&312.17&195.13&135.73\\ \hline
		Q4&208.51&142.12&136.69&130.51&125.60\\ \hline
		Q5&539.37&437.27&264.13&179.18&131.98\\ \hline
		Q6&220.29&133.08&108.56&77.02&67.42\\ \hline
		Q7&35.96&33.15&33.64&36.12&34.44\\ \hline
		Q9&575.29&310.14&297.75&190.08&145.75\\ \hline
		Q10&36.68&32.03&34.04&36.00&33.91\\ \hline
		Q11&580.72&412.59&300.78&183.67&138.56\\ \hline
		Q12&99.04&61.24&48.05&47.07&42.52\\ \hline
		Q13&575.09&402.15&291.11&186.86&120.88\\ \hline
		Q14&47.66&45.92&48.03&46.61&47.36\\ \hline
		Q15&36.90&32.04&36.53&36.06&35.46\\ \hline
		Q16&610.85&387.59&315.06&202.07&149.09\\ \hline
		Q17&NaN&NaN&NaN&NaN&NaN\\ \hline
		Q18&35.88&32.05&36.69&35.00&34.23\\ \hline
		Q19&597.83&499.11&305.14&196.61&140.85\\ \hline
		Q20&607.89&336.38&316.46&194.37&140.92\\ \hline
	\end{tabular}
\end{table}

\subsection{Loading and TeraSort} \label{appendix_load_terasort}
\begin{table}[H]
	\caption[Dataset Loading Time]{Dataset Loading Time (Minutes)}
	\centering
	\begin{tabular}{|l|r|r|r|r|r|}
		\hline
		&1 Node&2 Nodes&4 Nodes&8 Nodes&16 Nodes\\ \hline
		BSBM 10 Million &12.59&6.58&4.87&3.14&2.33\\ \hline
		BSBM 100 Million&117.66&66.03&33.05&26.90&20.28\\ \hline
		BSBM 1 Billion &1214.01&598.98&280.84&196.55&78.43\\ \hline
		DBPedia 150 Million&126.78&51.96&32.00&17.71&8.33\\ \hline
	\end{tabular}
\end{table}

\begin{table}[H]
	\caption{TeraSort Benchmark Results}
	\centering
	\begin{tabular}{|l|r|r|}
		\hline
		Step&Time (seconds)&Time (hours) \\ \hline
		TeraGen&3933.43&1.09 \\ \hline
		TeraSort&11234.88&3.12 \\ \hline
	\end{tabular}
\end{table}

TeraGen was run with the following command:
\footnotesize
\begin{verbatim}
hadoop jar hadoop-examples-1.0.3.jar teragen 10000000000 /terasort/input
\end{verbatim}
\normalsize
TeraSort was run with the following command:
\footnotesize
\begin{verbatim}
hadoop jar hadoop-examples-1.0.3.jar terasort /terasort/input /terasort/output
\end{verbatim}

\newpage
\section{Additional Figures}
\subsection{Hive Physical Query Plan: Example  (One Join)} \label{sec:appendix_physical_plan1}
\begin{figure}[h]
	\centering
	\includegraphics[width=0.635\textwidth]{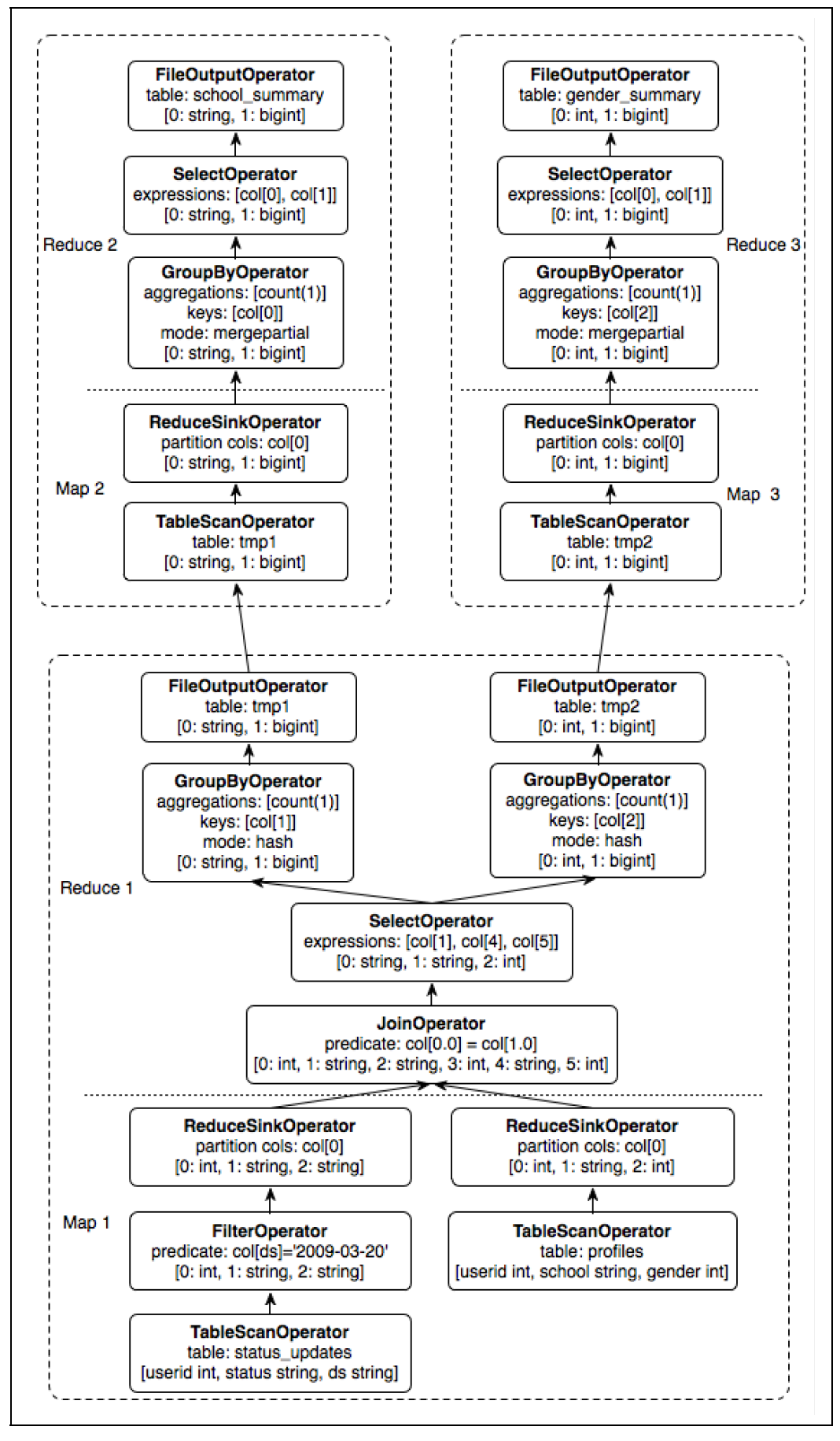}
\end{figure}

\newpage
\subsection{Hive Physical Query Plan: BSBM Query 1 (No Joins)} \label{sec:appendix_physical_plan2}

\begin{verbatim}
STAGE DEPENDENCIES:
Stage-1 is a root stage
Stage-2 depends on stages: Stage-1
Stage-0 is a root stage

STAGE PLANS:
Stage: Stage-1
Map Reduce
Alias -> Map Operator Tree:
TableScan
alias: rdf1
Filter Operator
Select Operator
expressions:
expr: key
type: string
expr: rdfs_label
type: string
outputColumnNames: key, rdfs_label
Group By Operator
bucketGroup: false
Reduce Output Operator
sort order: ++
Reduce Operator Tree:
Group By Operator
Select Operator
expressions:
expr: _col0
type: string
expr: _col1
type: string
outputColumnNames: _col0, _col1
File Output Operator

Stage: Stage-2
Map Reduce
Alias -> Map Operator Tree:
Reduce Output Operator
key expressions:
expr: _col1
type: string
sort order: +
Reduce Operator Tree:
Extract
Limit
File Output Operator
Stage: Stage-0
Fetch Operator
limit: 10
\end{verbatim}

\subsection{Hive Physical Query Plan: BSBM Query 5 (One Join)}
\begin{verbatim}
STAGE DEPENDENCIES:
Stage-6 is a root stage , consists of Stage-7, Stage-8, Stage-1
Stage-7 has a backup stage: Stage-1
Stage-4 depends on stages: Stage-7
Stage-2 depends on stages: Stage-1, Stage-4, Stage-5
Stage-8 has a backup stage: Stage-1
Stage-5 depends on stages: Stage-8
Stage-1
Stage-0 is a root stage

STAGE PLANS:
Stage: Stage-6
Conditional Operator

Stage: Stage-7
Map Reduce Local Work
Alias -> Map Local Tables:
Fetch Operator
Alias -> Map Local Operator Tree:
TableScan
HashTable Sink Operator

Stage: Stage-4
Map Reduce
Alias -> Map Operator Tree:
TableScan
Filter Operator
predicate:
expr: (bsbm_reviewfor = 'bsbm-inst_dataFromProducer21/Product978')
type: boolean
Map Join Operator
condition map:
Inner Join 0 to 1
Select Operator

Stage: Stage-2
Map Reduce
Alias -> Map Operator Tree:
Reduce Output Operator
Reduce Operator Tree:
Extract
Limit
File Output Operator

Stage: Stage-8
Map Reduce Local Work
Alias -> Map Local Tables:
Fetch Operator
Alias -> Map Local Operator Tree:
TableScan
alias: rdf1
Filter Operator
predicate:
expr: (bsbm_reviewfor = 'bsbm-inst_dataFromProducer21/Product978')
HashTable Sink Operator

Stage: Stage-5
Map Reduce
Alias -> Map Operator Tree:
TableScan
alias: rdf2
Map Join Operator
condition map:
Inner Join 0 to 1
handleSkewJoin: false
Select Operator

Stage: Stage-1
Map Reduce
Alias -> Map Operator Tree:
TableScan
alias: rdf1
Filter Operator
predicate:
expr: (bsbm_reviewfor = 'bsbm-inst_dataFromProducer21/Product978')
type: boolean
Reduce Output Operator
key expressions:
expr: rev_reviewer
type: string
Map-reduce partition columns:
expr: rev_reviewer
type: string
TableScan
alias: rdf2
Reduce Output Operator
Map-reduce partition
Reduce Operator Tree:
Join Operator
condition map:
Inner Join 0 to 1
handleSkewJoin: false
Select Operator
File Output Operator

Stage: Stage-0
Fetch Operator
limit: 20
\end{verbatim}


\end{document}